\documentstyle[a4p,11pt,epsfig,array,cite,pennames]{article}
\newcommand{\inmath}[1] {\ifmmode#1\else$#1$\fi}
\newcommand{\definmath}[2] {\def#1{\ifmmode#2\else$#2$\fi}}
\definmath{\PWpm} {\mathrm{W}^{\pm}}      
\definmath{\Pgtp} {\tau^{+}}        
\definmath{\Pgtm} {\tau^{-}}        
\definmath{\Pgtpm}   {\tau^{\pm}}         
\definmath{\Pgn}  {\nu}          
\definmath{\Pagn} {\overline{\nu}}     
\definmath{\Pq}      {\mathrm{q}}
\definmath{\Paq}  {\overline{\mathrm{q}}}
\definmath{\Pu}      {\mathrm{u}}
\definmath{\Pau}  {\overline{\mathrm{u}}}
\definmath{\Pd}      {\mathrm{d}}
\definmath{\Pad}  {\overline{\mathrm{d}}}
\definmath{\Ps}      {\mathrm{s}}
\definmath{\Pas}  {\overline{\mathrm{s}}}
\definmath{\Pc}      {\mathrm{c}}
\definmath{\Pac}  {\overline{\mathrm{c}}}
\definmath{\Pb}      {\mathrm{b}}
\definmath{\Pab}  {\overline{\mathrm{b}}}
\definmath{\Pt}      {\mathrm{t}}
\definmath{\Pat}  {\overline{\mathrm{t}}}
\definmath{\Pap}  {\overline{\mathrm{p}}}
\definmath{\Pan}  {\overline{\mathrm{n}}}
\definmath{\PaD}  {\overline{\mathrm{D}}}
\definmath{\PaDz} {\overline{\mathrm{D}}^{0}}
\definmath{\PaB}  {\overline{\mathrm{B}}}
\definmath{\PaBz} {\overline{\mathrm{B}}^{0}}
\definmath{\PsDpm}   {\mathrm{D}^{\pm}_{\mathrm{s}}}  
\definmath{\PcgLpm}  {\Lambda^{\pm}_{\mathrm{c}}}  
\definmath{\PD} {\mathrm{D}}     
\definmath{\PDst} {\mathrm{D}^{*}}     
\definmath{\PgLz} {\Lambda^{0}}        
\newcommand {\lsp}      {{{\tilde{\chi}}^{0}}_{1}}
\newcommand {\nln}      {{{\tilde{\chi}}^{0}}_{2}}
\newcommand {\Gravitino} {\tilde{\mathrm{G}}}
\newcommand {\nngg}       {\nu\overline{\nu}\gamma\gamma}
\newcommand {\nnggbra}       {\nu\overline{\nu}\gamma(\gamma)}
\newcommand {\nngggbra}       {\nu\overline{\nu}\gamma\gamma(\gamma)}
\newcommand {\ra}         {\rightarrow}
\newcommand {\ee}         {\mathrm{e}^+ \mathrm{e}^-}

\newcommand{\epem}   {\Pep\Pem}
\newcommand{\gamgam} {\Pgg\Pgg}
\newcommand{\mumu}   {\Pgmp\Pgmm}
\newcommand{\tautau} {\Pgtp\Pgtm}
\newcommand{\nunu}   {\Pgn\Pagn}

\newcommand{\eetogg}    {\epem\to\gamgam}

\newcommand{\eetomumu}     {\epem\to\mumu}
\newcommand{\eetotautau}   {\epem\to\tautau}

\newcommand{\eetonunu}     {\epem\to\nunu}

\newcommand{\BR}             {{\mathrm BR}}
\newcommand{\PX}             {{\mathrm X}}
\newcommand{\PY}             {{\mathrm{Y}}}
\newcommand{\eetonngg}       {\epem \to \nngg}
\newcommand{\eetonnggbra}    {\epem \to \nnggbra}
\newcommand{\eetonngggbra}   {\epem \to \nngggbra}
\newcommand{\eetoXY}         {\epem \to \PX\PY}
\newcommand{\eetoXX}     {\epem \to \PX\PX}
\newcommand{\XtoYg}       {\PX \to \PY\gamma} 

\newcommand{\sigXX}       {\sigma(\epem \to \PX\PX)}
\newcommand{\sigbrXY}    {\sigma(\eetoXY)\cdot\BR(\XtoYg)} 
\newcommand{\sigbrXX}    {\sigma(\eetoXX)\cdot\BR^2(\XtoYg)} 
\newcommand{\eetoXXs}   {\epem \to \nln \nln}
\newcommand{\eetoXYs}   {\epem \to \nln \lsp}
\newcommand{\XtoYgs}    {\nln \to \lsp\gamma}
\newcommand{\XtoYgg}    {\lsp \to \Gravitino\gamma}
\newcommand{\mx}         {M_{\PX}}
\newcommand{\my}         {M_{\PY}}
\newcommand{\myzero}        {\my\approx 0}
\newcommand{\betax}      {\beta_{\mathrm X}}
\newcommand{\mxmax}      {\mx^{\rm max}}
\newcommand{\mxmin}      {\mx^{\rm min}}
\newcommand{\roots} {\sqrt{s}}

\newcommand{\costhe} {\cos\theta}
\newcommand{\acosthe}   {|\costhe\,|}

\definmath{\GeV}  {\mathrm{GeV}}
\definmath{\GeVc} {\mathrm{GeV}\!/c}
\definmath{\GeVcc}   {\mathrm{GeV}\!/c^2}
\definmath{\MeV}  {\mathrm{MeV}}
\definmath{\MeVc} {\mathrm{MeV}\!/c}
\definmath{\MeVcc}   {\mathrm{MeV}\!/c^2}
\definmath{\MVm}  {\mathrm{MV}\!/\mathrm{m}}
\definmath{\keV}  {\mathrm{keV}}
\definmath{\keVcm}   {\mathrm{keV}\!/\mathrm{cm}}
\definmath{\kV}      {\mathrm{kV}}
\definmath{\km}      {\mathrm{km}}
\definmath{\meter}   {\mathrm{m}}
\definmath{\cm}      {\mathrm{cm}}
\definmath{\mm}      {\mathrm{mm}}
\definmath{\micron}  {\mu\mathrm{m}}
\definmath{\nm}      {\mathrm{nm}}
\definmath{\kg}      {\mathrm{kg}}
\definmath{\gram} {\mathrm{g}}
\definmath{\second}  {\mathrm{s}}
\definmath{\microsec}   {\mu\mathrm{s}}
\definmath{\degree}  {^\circ}
\definmath{\degC} {^\circ\mathrm{C}}
\definmath{\ohm}  {\Omega}
\definmath{\Mohm} {\mathrm{M}\Omega}
\definmath{\rad}  {\mathrm{rad}}
\definmath{\mrad} {\mathrm{mrad}}
\definmath{\nb}      {\mathrm{nb}}

\newcommand{\eqref}[1]  {(\ref{#1})}
\newcommand{\NIM} {Nucl.~Instrum.\ Methods}

\newcommand{\OPALColl}  {OPAL Collab.}

\newcolumntype{L} {>{$}l<{$}}
\newcolumntype{C} {>{$}c<{$}}
\newcolumntype{R} {>{$}r<{$}}

\begin{document}

\begin{titlepage}
\begin{center}{\large   EUROPEAN LABORATORY FOR PARTICLE PHYSICS
}\end{center}
\begin{flushright}
       CERN-PPE/97-132   \\ 6$^{\rm th}$ October 1997
\end{flushright}
\bigskip\bigskip\bigskip
\boldmath
\begin{center}{\huge\bf  Search for Anomalous Production of Photonic \\
Events with Missing Energy in $\epem$ Collisions \\
\vspace{1mm}
 at $\roots = 130 - 172~\GeV$
}\end{center}
\unboldmath
\bigskip\bigskip
\begin{center}{\LARGE The OPAL Collaboration
}\end{center}
\bigskip

\begin{abstract} 
Photonic events with large missing energy have been observed in 
$\mathrm e^+e^-$ collisions at centre-of-mass energies of 130, 136, 
161 and 172 GeV using the OPAL detector at LEP. Results are presented 
based on search topologies designed to select events with a single photon 
and missing transverse energy or events with a pair of acoplanar photons.
In both search topologies, cross-section measurements are performed within
the kinematic acceptance of the selection. These results are compared with the
expectations from the Standard Model processes $\eetonnggbra$ (single-photon)
and $\eetonngggbra$ (acoplanar-photons). No evidence is observed for 
new physics contributions to these final states. 
Upper limits on $\sigbrXY$ and $\sigbrXX$ \\ are derived for the case
of stable and invisible $\PY$. These limits apply to 
single and pair production of excited 
neutrinos ($\PX = \nu^*, \PY = \nu$), to neutralino production ($\PX=\nln, \PY=\lsp$), 
and to supersymmetric models in which $\PX = \lsp$ and $\PY=\Gravitino$ 
is a light gravitino. For the latter scenario, the results of the acoplanar-photons search are
used to provide model-dependent lower limits on the mass of the lightest neutralino.

\end{abstract}


\vspace{2.0cm}
\begin{center}
(To be submitted to Z. Phys. C)
\end{center}


\end{titlepage}

\begin{center}{\Large        The OPAL Collaboration
}\end{center}\bigskip
\begin{center}{
K.\thinspace Ackerstaff$^{  8}$,
G.\thinspace Alexander$^{ 23}$,
J.\thinspace Allison$^{ 16}$,
N.\thinspace Altekamp$^{  5}$,
K.J.\thinspace Anderson$^{  9}$,
S.\thinspace Anderson$^{ 12}$,
S.\thinspace Arcelli$^{  2}$,
S.\thinspace Asai$^{ 24}$,
S.F.\thinspace Ashby$^{  1}$,
D.\thinspace Axen$^{ 29}$
% ...
}\end{center}\bigskip
\bigskip
\vspace{-.5cm}
$^{  1}$School of Physics and Space Research, University of Birmingham,
Birmingham B15 2TT, UK
\newline
$^{  2}$Dipartimento di Fisica dell' Universit\`a di Bologna and INFN,
I-40126 Bologna, Italy
\newline
% ...
$^{  5}$Cavendish Laboratory, Cambridge CB3 0HE, UK
\newline
% ...
$^{  8}$CERN, European Organisation for Particle Physics,
CH-1211 Geneva 23, Switzerland
\newline
$^{  9}$Enrico Fermi Institute and Department of Physics,
University of Chicago, Chicago IL 60637, USA
\newline
% ...
$^{ 12}$Indiana University, Department of Physics,
Swain Hall West 117, Bloomington IN 47405, USA
\newline
% ...
$^{ 16}$Department of Physics, Schuster Laboratory, The University,
Manchester M13 9PL, UK
\newline
% ...
$^{ 23}$Department of Physics and Astronomy, Tel Aviv University,
Tel Aviv 69978, Israel
\newline
$^{ 24}$International Centre for Elementary Particle Physics and
Department of Physics, University of Tokyo, Tokyo 113, and
Kobe University, Kobe 657, Japan
\newline
% ...
$^{ 29}$University of British Columbia, Department of Physics,
Vancouver BC V6T 1Z1, Canada
\newline
% ...
\newline

\clearpage\newpage
\section{ Introduction }
\label{sec:intro}
 
This paper describes a search for photonic events with large missing
energy in $\epem$ collisions at centre-of-mass energies 
of 130, 136, 161 and 172 GeV. Two different search topologies are used. 
Cross-section measurements and search results from single-photon and acoplanar-photons 
topologies at $\roots=130$-$136$ GeV\cite{rf:OPALSP130} and $\roots=161$ 
GeV\cite{rf:OPALSP161} have 
been previously published. Those results have also been used to search for excited 
neutrinos with photonic decays at $\roots=130$-$136$~GeV~\cite{rf:excitedl} and 
$\roots=161$~GeV~\cite{rf:excited161}. In the analyses presented in this paper,
both the single and acoplanar-photons search techniques are based on those previously 
published by OPAL, but in each case the kinematic acceptance of the analysis has
been extended to lower energy and more forward angles, and the efficiency has
been increased by allowing for the possibility of photon conversions. These results
supersede our previous results.

The single-photon  and acoplanar-photons search topologies presented here
are designed to select events with one or more photons and significant missing 
transverse energy, indicating  the presence of at least one neutrino-like 
invisible particle which interacts only weakly with matter.
Results on photonic events without missing energy are presented 
in a separate paper.\cite{OPALgg}.


The single-photon search topology is sensitive to neutral events in which there are
one or two photons and missing energy, which within the Standard Model are expected 
from the $\eetonnggbra$ process. Measurements of 
single-photon production have been made in $\epem$ collisions at the $\PZz$ and at 
lower energies~\cite{rf:LEPSP,rf:OPALSP94,rf:lowe}. Results from centre-of-mass 
energies significantly above the $\PZz$ mass have also been 
reported~\cite{rf:OPALSP130,rf:LEPSP130}. The expected visible energies are 
sufficiently large at present centre-of-mass energies that doubly radiative neutrino
pair production can lead to two photons being detected; the experimental 
topology therefore includes such cases. 

The acoplanar-photons search topology is designed to select neutral events with 
two or more photons and significant missing transverse energy,
which within the Standard Model are expected 
from the $\eetonngggbra$ process.
The selection is 
designed to retain acceptance for events with a number of photons, $N_{\gamma}$, 
greater than two if the system formed by the three most energetic photons 
shows evidence for significant missing transverse energy.  

These photonic final-state topologies are sensitive to new physics of the 
type $\eetoXY$ and $\eetoXX$ where $\PX$ is neutral and decays radiatively 
($\XtoYg$) and $\PY$ is stable and only weakly interacting. For the general case of
massive $\PX$ and $\PY$ this includes conventional supersymmetric
processes\cite{Kane} $(\PX = \nln, \PY = \lsp)$. 
In this context it has been emphasised \cite{radN2} that the
radiative branching ratio of the $\nln$ may be large.
There is particularly good sensitivity
for the special case of $\myzero$ that applies both to the production of
excited neutrinos $(\PX = \nu^*, \PY = \nu)$ and to supersymmetric models
in which the lightest supersymmetric particle (LSP) is a light 
gravitino\footnote{The mass scale is typically $\cal{O}$(keV).},
and $\lsp$ is the next-to-lightest supersymmetric particle (NLSP) which decays 
to a gravitino and a photon,
($\PX=\lsp, \PY=\Gravitino$).
In this case, the branching ratio of this decay of the $\lsp$ is naturally large. 
Such a signature has been discussed in~\cite{ELLHAG} and more recently 
in~\cite{gravitinos,gravitinos2,LNZ} for a no-scale supergravity model and 
in~\cite{rf:chang} for a model with gauge-mediated supersymmetry breaking; 
in each case, $\lsp\lsp$ production cross-sections of order 1 pb are predicted 
at these centre-of-mass energies, for $M_{\lsp} \approx 50$~GeV.
Other types of new physics to which these search topologies are sensitive include
the production of invisible particles made visible through initial-state radiation and
the production of an invisible particle in association with a photon.
The acoplanar-photons search topology also has sensitivity to the
production of two particles, one invisible, or with an invisible decay mode, 
and the other decaying into two photons. 
Such events might arise from the production of a 
Higgs-like particle, $\rm S^0$ :
$\rm e^+e^-\rightarrow Z^0S^0$, followed by
S$^0$$\rightarrow \gamma\gamma$, $\rm Z^0\rightarrow \nu\overline{\nu}$.
Results for this model searching for the hadronic and leptonic $\rm Z^0$
decays appear elsewhere\cite{OPAL_Hgg}.

This paper will first briefly describe the detector, the data sample and the Monte 
Carlo samples used, including a discussion of event generators for 
$\eetonunu + n\gamma$. The event selection for each search topology will 
then be described, followed by cross-section measurements for 
$\eetonnggbra$ and $\eetonngggbra$ and comparisons with Standard Model 
expectations.
Implications of these results on the possibility of new physics processes of the type 
$\eetoXY$ or $\PX\PX$, $\XtoYg$ will be discussed. 

\section{Detector, Data Sample and Monte Carlo Samples}
\label{sec:detector}

The OPAL detector is described in detail elsewhere~\cite{rf:OPAL-detector}.
The measurements presented here are mainly based on the observation of 
clusters of energy deposited in the lead-glass electromagnetic 
calorimeters (ECAL) consisting of a 9,440 lead glass block array
in the barrel ($|\cos{\theta}| < 0.82$) with a 
quasi-pointing geometry, and two dome-shaped endcap arrays, each of 1,132
lead-glass blocks with axes coaxial with the beam axis
covering the polar angle range ($0.81 < |\cos{\theta}| < 0.984$).
In the overlap region, $0.785 < \acosthe < 0.815$, and at very forward angles,   
$\acosthe > 0.94$ the energy-resolution of the ECAL is slightly degraded 
relative to the nominal resolution. In some cases (where stated) these 
regions have been excluded from the analysis. 
These calorimeters, together with the gamma-catcher calorimeter, the forward 
calorimeters and the silicon-tungsten calorimeter (SiW),  provide a fully hermetic 
electromagnetic calorimeter down to polar angles of 33~mrad. The SiW 
calorimeter covers polar angles down to 24 mrad, however  the region around 
30 mrad lacks useful calorimetric coverage due to the installation, in 1996, 
of a thick tungsten shield designed to protect the tracking chambers from 
accelerator backgrounds while running at centre-of-mass energies well above the 
$\PZz$ resonance. The tracking system, consisting 
of a silicon microvertex detector, a vertex drift chamber (CV) and a large volume
jet drift chamber (CJ), is used to reject events  with prompt charged particles. 
The silicon microvertex detector consists of two concentric cylindrical layers of 
silicon microstrip arrays, each layer providing both an  azimuthal and longitudinal 
(along the beam direction) coordinate measurement. The two layer acceptance 
covers $|\cos{\theta}|<0.90$ for the 
161 and 172 GeV data-set while for the 
data acquired in 1995 at 130 and 136 GeV, the acceptance is limited
to $|\cos{\theta}|<0.75$.
Beam-related backgrounds and backgrounds arising from cosmic-ray interactions 
are rejected using  information from the time-of-flight system,
(TOF), the hadron calorimeter and muon detectors.

The data used in this analysis were recorded at $\epem$ centre-of-mass 
energies of 130.3, 136.2, 161.3, and 172.1~GeV, 
with integrated luminosities of 
2.30, 2.59, 9.89, and 10.28~pb$^{-1}$, respectively. The integrated 
luminosities are determined to better than 1\% from small-angle Bhabha
scattering events in the SiW luminosity calorimeter.
Triggers\cite{trigger} based on electromagnetic energy deposits in 
either the barrel or endcap electromagnetic calorimeters, and also  
on a coincidence of energy in the barrel electromagnetic calorimeter
and a hit in the TOF system, lead to 
full trigger efficiency for 
photonic events passing the event selection criteria described below.

For the expected Standard Model signal process,  
$\eetonunu + n\gamma$, the Monte Carlo generators 
NNGG03~\cite{rf:vvgmc}, NUNUGPV~\cite{rf:paviaMC} and KORALZ~\cite{rf:KORALZ}
were used. Modelling of these backgrounds is discussed in more detail in 
section~\ref{sec:evgens}.
For other expected Standard Model background processes, a number of different 
generators were used: RADCOR~\cite{rf:RADCOR} for $\epem \to \gamma \gamma (\gamma)$;
BHWIDE~\cite{rf:BHWIDE} and TEEGG~\cite{rf:TEEGG} for $\epem \to \epem (\gamma)$;
and KORALZ for  $\eetomumu(\gamma)$ and $\eetotautau(\gamma)$.

To simulate possible new physics processes of the type  $\eetoXY$ 
and $\eetoXX$ where $\PX$ decays to $\PY\gamma$ and 
$\PY$ escapes detection, the SUSYGEN~\cite{rf:SUSYGEN} Monte Carlo 
generator was used to produce neutralino pair events of the type 
$\eetoXYs$ and  $\eetoXXs$, $\XtoYgs$,  
with isotropic angular
distributions for the production and decay of $\nln$ and 
including initial-state radiation.
SUSYGEN Monte Carlo events were generated at each centre-of-mass energy, for 16-24
points in the kinematically accessible region of the 
($\mx$, $\my$) plane for which $\mx-\my\ge 5$ GeV, depending on the centre-of-mass energy.
For the case $\myzero$, the efficiencies for $\PX\PY$ and $\PX\PX$ production 
obtained from these Monte Carlo samples 
are consistent 
within statistical errors with the efficiencies obtained from the OPAL 
$\nu^*\bar{\nu}$ and $\nu^*\bar{\nu}^*$ excited neutrino Monte Carlo samples 
respectively\cite{rf:excitedl}.
All the Monte Carlo samples described above were processed through the
OPAL detector simulation~\cite{rf:GOPAL}.

\boldmath
\subsection{Event Generators and Analytical Calculations of
$\eetonunu + n\gamma$}
\unboldmath
\label{sec:evgens}
The present status of event generators for the 
Standard Model process $\eetonunu + n\gamma$, $n\ge 1$, is very unsatisfactory for
the centre-of-mass energy region, 130 GeV $<\sqrt{s}<$ 172 GeV,  relevant to the analyses
presented here.

For $\sqrt{s} \approx M_{\rm Z}$, two event generators, NNGG03 and 
KORALZ were used extensively for
studies of $\eetonnggbra$
with a demonstrated agreement\cite{rf:colas} between the cross-section predictions 
of better than  1\%.
NNGG03 is  designed for $\eetonnggbra$ at
$\sqrt{s} \approx M_{\rm Z}$ with inclusive exponentiation 
of soft photons and the hard photon matrix element for
$\eetonngggbra$ for the Z exchange diagrams only.
At higher centre-of-mass energies,
it has not been maintained officially by the authors, nor 
claimed to be reliable.
The absence of a complete lowest order calculation 
for $\eetonngg$ and higher order corrections ($\eetonngggbra$)
make it necessarily incomplete for
$\eetonngggbra$.
The KORALZ event generator, primarily designed for 
$\ee \ra \mumu (\gamma)$ and 
$\ee \ra \tautau (\gamma)$
at $\sqrt{s} \approx M_Z$,
can also generate $\eetonunu + n\gamma$, $n\ge 0$,
using the Yennie-Frautschi-Suura
approach\cite{rf:YFS}
to 
explicitly generate an arbitrary number of 
additional initial state photons.
This generator is maintained by the authors
for $\sqrt{s} \gg M_{\rm Z}$, but no specific 
publications exist yet attesting to its 
accuracy for either $\nnggbra$ or $\nngggbra$ 
in this regime.
The NUNUGPV analytical calculation is designed for
$\sqrt{s} \approx M_{\rm Z}$ and 
$\sqrt{s} \gg M_{\rm Z}$ using the $p_T$
dependent structure function approach
to estimate $\eetonnggbra$ with a claimed accuracy
of 1-2\% for $ 150 < \sqrt{s} < 175 $ GeV.
An event generator based on this calculation is 
also available which includes the emission of 
an additional photon from each beam.
This feature is designed to permit estimation of the
effect of $\nngggbra$ events on the $\nnggbra$ acceptance.
It is not intended as an accurate estimate of the 
$\nngggbra$ cross-section.
In a previous publication\cite{rf:OPALSP161}, we
used this feature inappropriately to
estimate the expected contribution from
$\eetonngggbra$.

Recently, independent calculations have been made of the $\eetonngg$ (lowest order)
cross-section  using CompHep\cite{rf:comphep} 
(by Ambrosanio\cite{rf:Sandro}), and 
using HELAS\cite{rf:helas} (by Mrenna\cite{rf:Mrenna}).
Calculations of
the $\eetonngggbra$ cross-section by Mrenna, and 
by Bain and Pain \cite{rf:bread-bath}
using GRACE\cite{rf:grace} and CompHep have also been made.
These are approximately a factor of two lower than the 
$\nngggbra$ cross-section predictions we obtain using the NNGG03 and NUNUGPV event
generators. The estimated $\nngggbra$ cross-section from KORALZ
agrees reasonably well with the independent calculations.
For $\ee \ra \nnggbra$, we have found that the
estimated cross-section from KORALZ is lower by about 10\% compared with 
NUNUGPV within the kinematic acceptance of the single-photon
selection, described in section~\ref{sec:intro}.

KORALZ is used to estimate the detection efficiency of
$\eetonnggbra$ and $\eetonngggbra$ given its more 
complete treatment of 
events with multiple photons. Generator studies indicate that it also provides a
reasonable estimate of the fraction of two photon events ($\eetonngggbra$).
The estimated efficiencies obtained using KORALZ are compared with those
obtained using NNGG03 and NUNUGPV and only small
differences are found, indicating that the experimentally measured
cross-sections are relatively insensitive to the choice of generator.

For coherence in the comparisons of data with Monte Carlo, we use
KORALZ. In calculating upper limits on new processes for the 
single-photon topology, despite the claimed 1-2\% accuracy of NUNUGPV, 
we use the background estimate from KORALZ 
which is the lower of the two and is therefore expected to be conservative.
Given the current status of 
calculations of $\eetonngggbra$, where factor of two differences between 
some cross-section estimates are not yet fully understood, 
in calculating limits on new processes for the 
acoplanar-photons topology we do not take into account the
$\nngggbra$ background estimate.

\section{Photonic Event Selection}
\label{sec:selection}

This section describes the criteria for selecting single-photon and 
acoplanar-photons events.
The kinematic acceptance of each topology is defined in terms of the photon energy, 
$E_{\gamma}$
and the photon polar angle, 
$\theta$, defined with respect to the electron beam direction.
In order to simultaneously maintain sensitivity to low photon energies and
to retain acceptance at high polar angles, the kinematic acceptance of each
selection is composed of two (possibly overlapping) parts:
\begin{description}
\item[Single-Photon -]
One or two photons accompanied by invisible particle(s):
\begin{itemize}
\item At least one photon with $x_{T} > 0.05 $ and $\acosthe < 0.82$, or,  
\item At least one photon with $x_{T} > 0.1 $ and ($0.82 < \acosthe < 0.966)$.
\end{itemize}
\item[Acoplanar-Photons -] 
Two or more photons accompanied by invisible particle(s):
\begin{itemize}
\item At least two photons with $x_{\gamma}>0.05$ and $15^{\circ}<\theta<165^{\circ}$, or,
\item One photon with $E_{\gamma} > 1.75$ GeV  and $\acosthe < 0.8$ and a second photon \\
satisfying $E_{\gamma} > 1.75$ GeV and $15^{\circ}<\theta<165^{\circ}$ ($\acosthe < 0.966)$.
\end{itemize}
\end{description}
where the scaled energy, $x_{\gamma}$, is defined as
$E_{\gamma}/E_{\mathrm{beam}}$, and the scaled transverse
momentum,
$x_{T}$, is defined as $E_{\gamma}\sin{\theta}/E_{\mathrm{beam}}$.

In each of the topologies, it is desirable to retain acceptance for events 
with an additional photon, if the resulting multi-photon system is
still consistent with the presence in the event of significant 
missing energy. This reduces the sensitivity of each measurement to the
modelling of higher-order contributions.

\subsection{Single-Photon Event Selection Description}
\label{sec:g1_selection}

Events pass the single-photon selection if they satisfy the 
criteria listed below. These selection criteria are similar 
to previous OPAL analyses of photonic events with missing 
energy but have increased acceptance and efficiency: 

\begin{itemize}
\item{
{\bf Angular acceptance and minimum transverse energy.}
An event is considered to contain a photon candidate if the primary
electromagnetic cluster (that with the highest deposited energy in 
the barrel or endcap calorimeters) is in the region 
$15^\circ < \theta < 165^\circ$
($\acosthe < 0.966$) and has a scaled transverse energy, $x_{T}$, 
that exceeds 0.1. Events with a primary cluster having $0.05 < x_{T} < 0.1$
and in the barrel region $\acosthe < 0.82$ are also accepted
if they have associated TOF information with good timing, as described 
in the fourth selection criterion below.
Events are considered to have more than one photon if additional
electromagnetic clusters are found in the barrel or endcap
calorimeter ($\acosthe < 0.984$) having deposited energy
exceeding 300~MeV. 
}
\item{
{\bf Charged track veto or photon conversion consistency requirements.}
Events are vetoed if there is a charged track with ten or more hits in 
the central detector, unless the track is consistent with arising from a 
photon conversion.
Events having no charged tracks are called non-conversion
candidates. Photon conversion consistency requires that the
primary photon candidate cluster be associated 
with the highest $p_T$ track in the event within 100 mrad 
in both azimuthal and polar angle. This
track should not be prompt, i.e. the cluster is not accepted as a
possible photon conversion if there are two or more 
(out of a maximum possible of four) associated silicon
microvertex detector hits for photons within its two-layer
acceptance,
or a minimum number of hits in the CV inner axial wires 
(for photon polar angles beyond the microvertex two-layer acceptance)
that are associated in azimuth to the above cluster. 
Events for which  the primary photon is consistent with a photon
conversion are called conversion candidates. 
Events with conversion candidates are rejected if they have at
least 2 tracks, reconstructed from axial-wire hits in CV, 
with an opening angle in the transverse
plane exceeding 45 degrees. This cut removes most of the events
having charged tracks which do not arise from a single photon conversion. }
\item{
{\bf Cluster extent.} Only clusters containing more than one ECAL block are considered
as photon candidates. The primary electromagnetic cluster, combined 
with any clusters contiguous with it, must be consistent with the cluster size and 
energy sharing of blocks for a photon coming from near the interaction
point. The cluster size varies in both azimuthal and polar angle
extent as a function of $\acosthe$. The cluster extent cuts
are parametrized in $\acosthe$ accordingly.
}   
\item{
{\bf Forward energy vetoes.}
The total energy deposited in each of the left and right
forward calorimeters and in each of the left and right 
SiW calorimeters
must be less than 5~GeV.
The most energetic gamma-catcher cluster must have an energy
of less than 5~GeV. These vetoes serve to ensure 
that photon candidate events are not accompanied by any
event activity in the forward regions.}  
\item{
{\bf Muon veto.}
Events are rejected if there are any muon track segments
reconstructed in the barrel or endcap muon chambers, or 
in the barrel, endcap or pole-tip hadron calorimeters.
Events are also rejected if three or more
of the outer eight layers of the barrel hadron calorimeter 
have strips hit in any $45^{\circ}$ azimuthal octant. 
The muon veto is used primarily to remove cosmic ray background.
}
\item{
{\bf Timing measurement in TOF system for low $x_{T}$ and
conversion candidates.}
An electromagnetic cluster is said to have an associated TOF hit
if it is matched within 50~mrad in azimuthal angle by a good quality TOF 
counter signal produced by the photon converting 
before or in the coil which is located in front of the TOF.
A cluster with an associated TOF hit has good timing if
the measured arrival time of the photon at the TOF
is within 20 ns of the expected time for a photon 
originating from the interaction point. 
For all events with a photon conversion candidate in the
barrel region $\acosthe < 0.82$ and for events
with a low $x_{T}$ ($0.05 < x_{T} < 0.1$) non-conversion candidate 
in the barrel region, we require an associated TOF hit with good timing.
For all other events with the primary photon
in the barrel region, if there is an associated TOF hit, it must have good timing.}
\item{
{\bf Special background vetoes for events with no TOF hit.}
If the primary non-conversion candidate
photon has no associated TOF hit, three different background vetoes
are applied. The first rejects events in which any of the three
muon triggers\cite{trigger} (barrel and two endcaps) were present. This veto
removes much of the cosmic ray background. The second
looks for a series of calorimeter clusters at the same $r$ and $\phi$
as the primary cluster, 
but at different $z$. This veto
rejects beam halo type backgrounds. The third looks for
a series of hits in the outer layers of the hadron calorimeter.
This veto rejects both cosmic rays and beam related backgrounds.}
\end{itemize}

Events with a second photon candidate are rejected as background from 
$\Pep\Pem \to \Pgg\Pgg(\Pgg)$ whilst retaining acceptance for events 
with two photons and missing energy
if any of the following criteria are satisfied:
\begin{itemize}
\item{The total energy of
the two clusters exceeds $0.9\roots$.}
\item{The acoplanarity angle\footnote{Defined as $180^{\circ}$ minus the 
opening angle in the transverse plane.}
of the two clusters is less than $2.5^\circ$.}
\item{The missing momentum vector 
calculated from the two clusters satisfies
$|\cos{\theta_{\mathrm{miss}}}| > 0.9$.}
\item{A third electromagnetic cluster is detected with
deposited energy exceeding 300~MeV.}
\item{The transverse momentum of the two photon system does not
exceed 0.05 of the beam energy.}
\item{For events with at least one of the two clusters in the region
$\acosthe > 0.95$, the variable $b_{T}$ is less than
0.1, where $b_{T} = (\sin{\theta}_1 + \sin{\theta}_2) |\cos\left[(\phi_1 -
\phi_2)/2\right]|$. This amounts to a stronger acoplanarity cut for events
with at least one forward photon.}
\end{itemize}

\begin{figure}[b]
\centerline{\epsfig{file=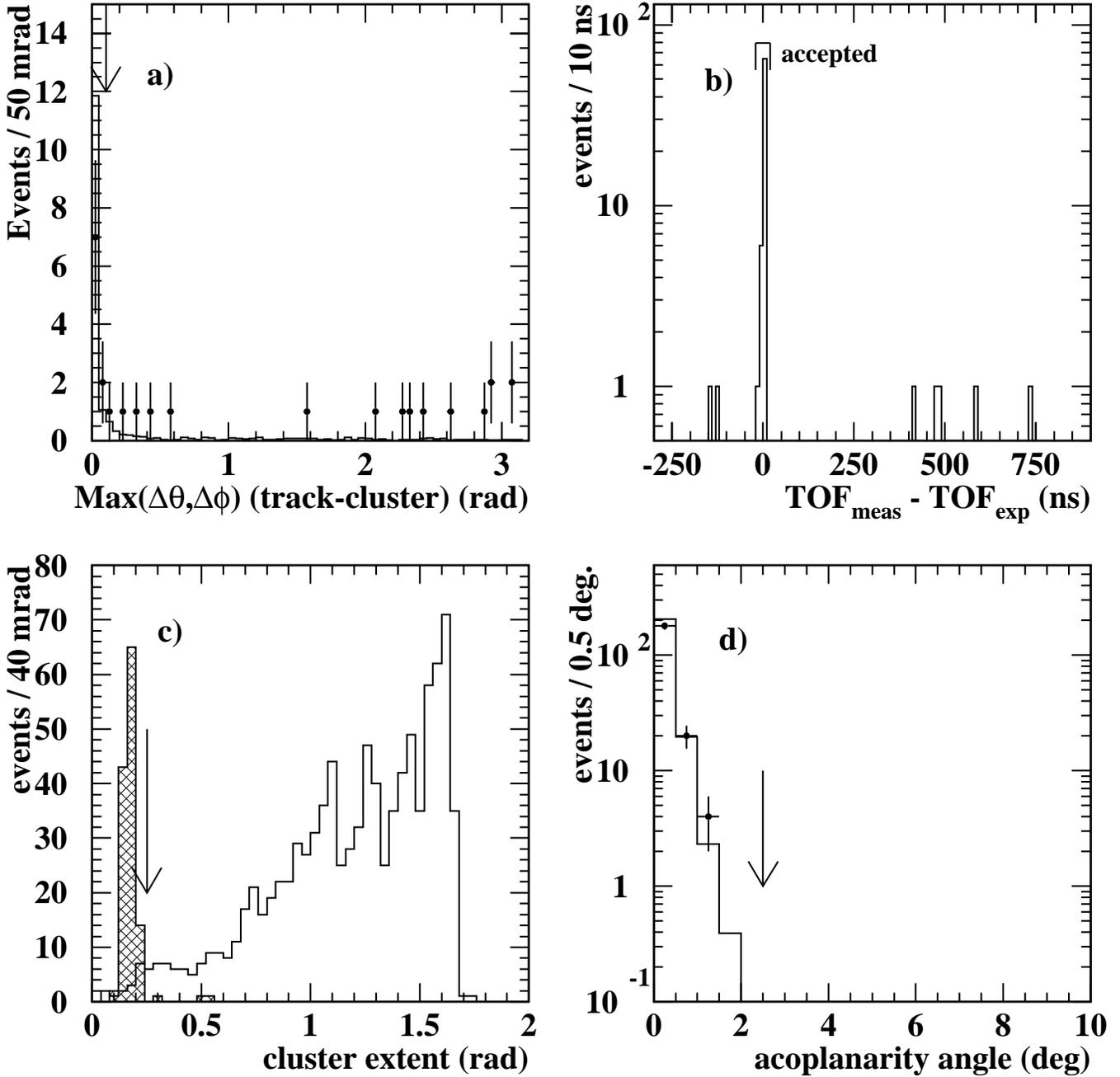,width=11cm,
bbllx=125pt,bblly=150pt,bburx=430pt,bbury=680pt}}
\vspace{-.5cm}
\caption{ 
For single-photon conversion candidates, a) shows the maximum
angular separation in $\theta$ and $\phi$ of the primary photon candidate
and the highest $p_T$ track in the event. For the non-conversion single-photon
selection, b) shows the difference between the observed TOF timing and the 
timing expected for a photon from the interaction point, for data events
passing all cuts or failing only the timing cut. For the acoplanar-photons
selection, c) shows the maximum cluster extent for data events failing only the 
anti-$\gamma\gamma(\gamma)$ cuts (shaded histogram) and for data 
events failing the TOF requirements or the TOF requirement and the special
background vetoes (unshaded histogram).
For the acoplanar-photons selection, d) shows the distribution of the 
acoplanarity angle for data events failing only the total energy cut and/or the cut
on $p_T(\gamma\gamma)$ (predominantly $\eetogg$).
In a) and d) the solid points with error bars show the OPAL data while the overlaid
histograms represent the expectation from $\nnggbra$ and $\eetogg (\gamma)$ Monte Carlo,
respectively, normalized to the luminosity of the data. 
}
\label{fig:cuts}
\end{figure}

For the conversion selection, Figure~\ref{fig:cuts} a) shows the maximum 
of the angular separation, in $\theta$ and $\phi$, of the primary
photon candidate and the highest $p_T$ track in the event. Overlaid as
a histogram is the expected distribution from $\nnggbra$ Monte Carlo, normalized to the
integrated luminosity of the OPAL data. The cut at 100 mrad rejects contributions
from cosmic rays. For the non-conversion selection,
Figure~\ref{fig:cuts} b) shows the difference between the measured TOF timing
and that expected for a photon from the interaction point, for events passing
all selection criteria or failing only the TOF timing requirement. The seven
events outside the accepted region of $\pm{20}$ns are rejected as cosmics.


\subsection{Acoplanar-Photons Event Selection Description}
\label{sec:g2_selection}

The acoplanar-photons selection has two overlapping regions of kinematic acceptance,
in order to retain both sensitivity to low-energy photons and acceptance at large
$\acosthe$. These selections are based on analyses previously published by OPAL
using data collected at centre-of-mass energies of 130-136 GeV\cite{rf:OPALSP130}. 
The analysis presented in this paper has increased acceptance and efficiency relative 
to the previous OPAL analyses. The selection criteria are summarized below:

\begin{itemize}
\item{
{\bf Angular acceptance and minimum energy.}
Events are accepted as candidates if there are at least two electromagnetic clusters
with scaled energy, $x_{\gamma}$, exceeding 0.05 in in the polar-angle region
$15^{\circ} < \theta < 165^{\circ}$ ($\acosthe < 0.966$). In order to retain 
sensitivity to physics processes producing low-energy photons, the minimum
energy requirement is relaxed to 1.5 GeV deposited energy for photon candidates
in the polar-angle region $\acosthe < 0.8$.
This energy deposition corresponds to a photon with an energy of about 
1.75 GeV\cite{rf:OPALSP94}. 
These two selections are referred to below as the ``standard'' and 
``low-energy'' selections, respectively. Background vetoes are applied
differently for the two parts of the selection, as described below.
}
\item{
{\bf Charged track veto or photon conversion consistency requirements.}
We use selection criteria designed to reject events having
tracking information consistent with the presence of at least
one charged particle originating from the interaction point.
These criteria are designed to retain acceptance for events in which 
one or both of the photons convert. For the standard selection 
we use hit information from the central jet-chamber, the vertex drift chamber, 
and the silicon microvertex detector (for $\acosthe < 0.75$ (0.9) for data taken at 
$\roots =$ 130-136 (161, 172) GeV).
These three detectors form independent estimators for the existence of
charged particle activity. Events in which charged particle activity 
is associated in azimuth with both photon candidates are rejected unless the 
signal is from the jet chamber only or from the microvertex detector only. 
Events in which only one photon candidate has azimuthally associated charged particle 
activity  are rejected if all (two or three) layers of charged particle detection 
registered activity. To address possible backgrounds from
$\rm e^+e^-\rightarrow\ell^+\ell^-\gamma\gamma$, an additional veto requires that 
there be no reconstructed charged track with transverse momentum exceeding 1~GeV,
with associated hits in the axial layers of the vertex chamber,  
and separated from each of the photon candidates by more than $15^\circ$. 

The low energy part of the selection requires that there be no 
reconstructed charged track in the event with 20 or more reconstructed hits in the 
central jet-chamber.

}
\item{
{\bf Cluster extent.} Any photon candidate within the polar angle region 
$\acosthe < 0.75$ is required to have an angular cluster extent that is less
than 250 mrad in both $\theta$ and $\phi$.
}
\item{
{\bf Forward energy vetoes.}
The forward vetoes described for the single-photon analysis
are applied with the same thresholds.
}
\item{
{\bf Muon veto.}
To suppress backgrounds arising from cosmic-ray muon 
interactions or beam halo muons which can deposit significant 
energy in the calorimeter, the events must pass the muon veto 
described  for the single-photon analysis. Additionally
the special background vetoes described for the single-photon selection are 
applied to events in which no TOF information is present. 
}
\item{
{\bf Timing measurement in TOF system.}
Requirements on time-of-flight (TOF) information are defined separately for the
two parts of the kinematic selections. 
For the low-energy part of the selection, we require that the photon in the 
barrel region has an associated TOF hit with good timing (as defined for the 
single-photon analysis in section~\ref{sec:g2_selection}).
For the standard selection, for events in which both of the photon candidates 
lie within $\acosthe < 0.82$, at least one of them must have an
associated TOF hit with good timing. For all events we
reject the event if either of the photon candidates has an associated TOF 
hit with bad timing. Finally, if there is a charged track associated with a 
cluster within the polar 
region $\acosthe < 0.82$, the requirement of an associated TOF hit 
with good timing is applied.  
}

\end{itemize}
Acoplanar photons events can be faked by cosmic-ray and beam-halo events
in which a muon grazes the electromagnetic calorimeter. Such events can
produce large clusters which are split by the clustering algorithm to produce
two or more clusters. Since it is difficult to model such backgrounds it is 
useful to have a great deal of redundancy in the procedures used to reject
these contributions. This redundancy is provided by the selection criteria
outlined above. Figure~\ref{fig:cuts} c) shows the maximum cluster
extent for events in which both photons are in the polar-angle region
$\acosthe < 0.75$, where cluster-extent cuts are applied. The shaded 
histogram shows the cluster-extent distribution for real photons
coming from the interaction point. These come dominantly from 
$\eetogg$ events  selected by removing the anti-$\gamma\gamma (\gamma)$ cuts
(described below).
The shaded histogram shows the same distribution for events failing the
TOF cuts and (possibly) the special background vetoes. The cut at 250 mrad
is indicated by the arrow.
Additional suppression of such events, as well as of beam-wall/beam-gas events and
instrumental backgrounds in the overlap and endcap regions, is obtained by 
imposing the following requirements:
\begin{itemize}
\item Events are vetoed if there is a reconstructed charged track 
with at least 20 hits in the jet chamber and a 
$|z_0|$ larger than 50 cm, where $z_0$ is the $z$ coordinate 
of the point of closest approach of the track to the beamline in the transverse plane. 
\item Events are vetoed if the total number of ECAL clusters
having more than 1 GeV of deposited energy is larger than five.
\item If both photons have $\acosthe > 0.75$, the opening angle 
$\psi$ of the two-photon system is required to satisfy ${\rm cos}\psi<0.98$. 
Otherwise the requirement is that the azimuthal separation of the two 
candidate clusters be greater than $2.5^\circ$.  
\end{itemize}

Finally, background from $\Pep\Pem \to \Pgg\Pgg(\Pgg)$ is rejected,
whilst retaining acceptance for the signal topology, 
if any of the following 
criteria are satisfied:
\begin{itemize}
\item The total visible energy of the event exceeds $0.95\roots$.
\item The acoplanarity angle of the two highest-energy 
clusters is less than $2.5^{\circ}$.
\item The missing momentum vector calculated from the two highest-energy photon
candidates satisfies $\mathrm |\cos{{\theta}_{miss }}|~>~0.95$.
\item The transverse momentum of the two-photon system is less than 
0.05$E_{\rm beam}$;
events having a third photon candidate (with $E_{\gamma}>300$ MeV) 
are rejected unless the three photon system is significantly aplanar 
(sum of the three opening angles $< 350^{\circ}$) and 
the transverse momentum of the three-photon system exceeds 0.1$E_{\rm beam}$.
\end{itemize}

Figure~\ref{fig:cuts} d) shows the distribution of the acoplanarity angle for 
events passing all cuts or failing one or both of the total-energy cut and 
the cut on the scaled transverse momentum of the two-photon system. 
The OPAL data is shown as solid points with error bars.
Overlaid as a histogram is the expected distribution, from $\eetogg (\gamma)$ Monte 
Carlo, normalized to the luminosity of the OPAL data. The cut at 2.5$^{\circ}$ 
is indicated.

\section{Results}
\label{sec:results}
The results of the single-photon 
and acoplanar-photons selections
are given below in sections~\ref{sec:g1_results}
and section~\ref{sec:g2_results} respectively.
In each section, the measured cross-sections for the search topology are given 
and compared with Standard Model expectations, and
the results of the XY and XX searches are then described.
Upper limits on
$\sigbrXY$ and $\sigbrXX$
are given, respectively.
This is done both 
for the general case of massive $\PX$ and $\PY$, applicable to conventional
supersymmetric models in which 
$\PX = \nln$ and  $\PY = \lsp$, 
and also separately for the special case of $\myzero$, which applies 
both to
single and pair production of neutralinos 
in  supersymmetric models in which the LSP is a light gravitino and to
single and pair production of excited neutrinos.
These results are used to set limits on the production of excited neutrinos ($\nu^*$) 
in a separate paper~\cite{ref:opal_exlp_172}.
For all such limits, the efficiencies were evaluated with
the decay length of $\PX$ set to zero. The efficiencies are 
unaffected if the decay length is much less than the 
distance from the interaction point to the electromagnetic calorimeters ($\approx$2 m).

For the single-photon and acoplanar-photons analyses, XY and XX Monte Carlo
events were generated at each centre-of-mass energy for a variety of mass 
points in the kinematically accessible region of the $(\mx,\my)$ plane.
To set limits for arbitrary $\mx$ and $\my$,
the efficiency over the entire $(\mx,\my)$ plane
is parametrized using the 
efficiencies calculated at the 
generated mass points.
For the single-photon search topology, the region with $\mx+\my < M_{\rm Z}$
is kinematically accessible at $\sqrt{s} \approx M_{{\rm Z}}$
and strong limits have already been 
reported\cite{rf:LEP1XY}. For the acoplanar-photons search topology,
we restrict the search to $\mx$ values larger than about $M_{\rm Z}$/2.
At lower masses, limits have been reported at $\sqrt{s} \approx M_{{\rm Z}}$
\cite{LEP1XX} and
possible radiative return to the $\rm Z$ followed by $\rm Z\to XX$
would yield very different event kinematics than those of the signal Monte
Carlo events used for this study. For both search topologies,  at
values of $\mx-\my<5$~GeV, the estimated efficiency decreases significantly due 
to event kinematics that yield low photon energies. For that reason no limits
are set in this region.

\boldmath
\subsection{Single-Photon Results $\gamma(\gamma)+{E_T\hspace*{-1.0em}/\quad}$ }
\unboldmath
\label{sec:g1_results}

After applying the selection criteria of the single-photon
selection to the $\roots = 130$-$172 $ GeV data samples,
a total of 138 events are selected. 
The expected non-physics background is $2.3 \pm 1.1$ events, consisting
solely of cosmic ray and beam related backgrounds. 
This non-physics background has been estimated from events 
detected out of time and using a visual scan with looser cuts.
The expected physics backgrounds from plausible sources,
$\eetogg(\gamma)$, Bhabha events with initial or
final-state radiation and $\ee \ra \mumu \gamma$ and
$\ee \ra \tautau \gamma$, have
been evaluated to be less than 0.4 events 
at 95\% confidence level (CL) and are therefore
considered to be negligible for the cross-section measurement.
For each of the four centre-of-mass energies,
Table 1 shows the number of events observed, 
the number of events expected from the Standard Model process 
$\eetonunu \gamma(\gamma)$ evaluated using
the KORALZ generator, the 
NNGG03 generator ($\roots$ = 130 and 136 GeV) and the 
NUNUGPV generator ($\roots$ = 161 and 172 GeV), and 
the number of background events expected from non-physics processes.
The estimated efficiency
for selecting $\eetonnggbra$ events within the
kinematic acceptance of the single-photon selection
is also given, as is the corresponding 
measured $\eetonnggbra$ cross-section
within this kinematic acceptance, corrected for detector and selection
efficiencies, and subtracting the estimated non-physics background.
For both the single-photon and acoplanar-photons selections,
efficiency losses due to detector occupancy range from about 
(3-5)\% at the different centre-of-mass energies. Here and elsewhere in this 
paper, unless otherwise stated, quoted efficiencies include these losses 
and those due to detector status requirements. 

The number of events observed
agrees with the number of events expected from $\eetonnggbra$
plus the background. The two Monte Carlo generators give similar
results although the KORALZ generator has a systematically lower
cross-section than NNGG03/NUNUGPV. Following the discussion in 
section 2, the KORALZ Monte Carlo sample is
used for all subsequent measurements and results 
concerning the $\eetonnggbra$ process unless 
explicitly mentioned otherwise.

\begin{figure}[b]
\centerline{\epsfig{file=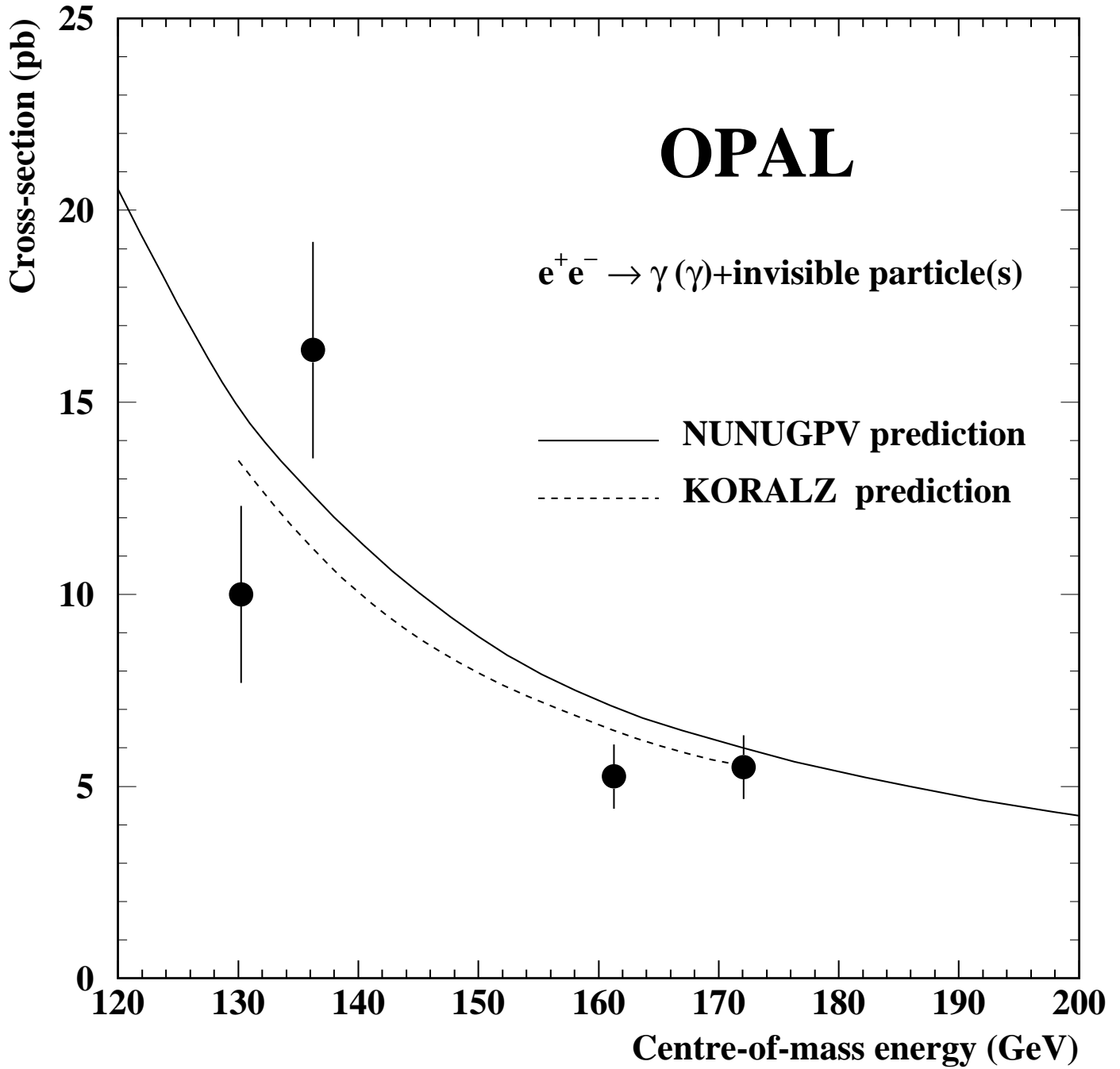,width=11cm,
bbllx=125pt,bblly=150pt,bburx=430pt,bbury=680pt}}
\caption{ The measured value of $\sigma(\epem \to \gamma (\gamma)$ + invisible particle(s)),
within the kinematic acceptance of the single-photon selection,
as a function of $\roots$.
The data points with error bars are OPAL
measurements at centre-of-mass
energies of 130, 136, 161 and 172 GeV. 
The curves are the predictions for the
Standard Model process $\eetonnggbra$ from the KORALZ generator and the
NUNUGPV analytical calculation.
}
\label{f:cross}
\end{figure}

Systematic errors on the cross-section measurement
arising from uncertainties on the electromagnetic calorimeter energy scale and
resolution, the description of the detector material and consequent conversion 
probabilities of photons in the central detector volume and coil, the 
integrated luminosity
measurement, and the detector occupancy estimate, 
have been considered and evaluated to be negligible
with respect to the statistical error.
A relative systematic error of 4\% is assigned to the cross-section measurement.
This uncertainty comes dominantly from the estimated
uncertainty on the efficiency based on comparing the different event generators. 
The modelling of the $\ee \ra \nnggbra$ event fraction with a 
second photon detected in the forward detectors ($|\cos{\theta}|>0.984$), 
and therefore rejected by the forward energy vetoes is 
expected to be the main reason for the 
observed efficiency differences reported in Table 1.
The cross-sections as a function of centre-of-mass
energy are plotted in Figure~\ref{f:cross}. The curves show the
predicted cross-sections
from the KORALZ event generator and 
the NUNUGPV analytical calculation 
for the Standard Model process
$\eetonnggbra$. The data are generally consistent with the predictions
but do not favour either estimate.

\begin{figure}[b]
\centerline{\epsfig{file=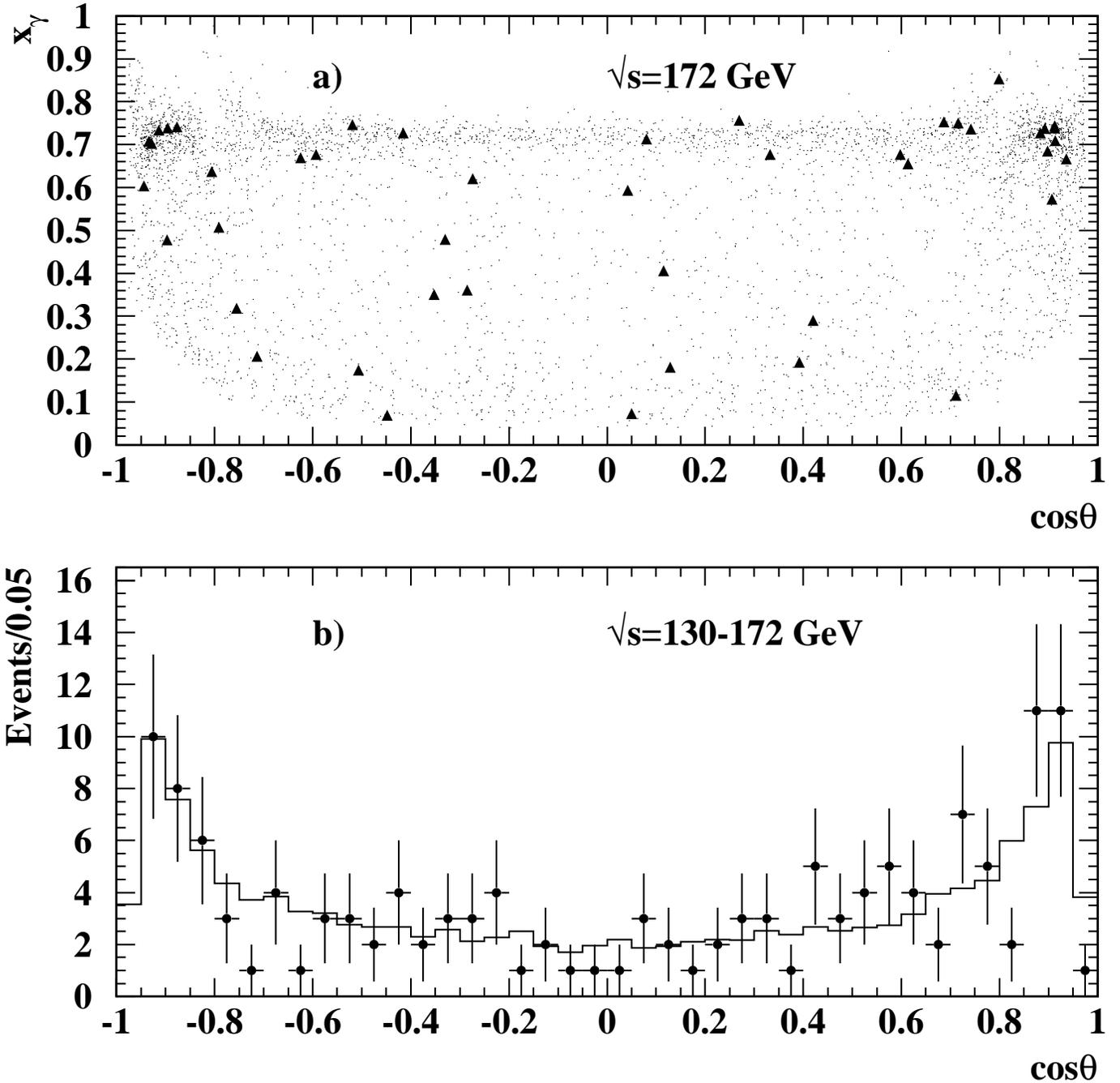,width=11cm,
bbllx=125pt,bblly=150pt,bburx=430pt,bbury=680pt}}
\caption{ a)~Distribution of $x_{\gamma}$ vs $\cos\theta$
for the most energetic photon in the single photon selection. 
The fine points are the KORALZ $\eetonnggbra$ Monte Carlo
and the solid triangles are the data. This plot is for
$\roots$ = 172 GeV.
b)~The $\cos\theta$ distribution for 
the most energetic photon in the single photon selection. 
The points with error bars are the data and the histogram is the
expectation from the KORALZ $\eetonnggbra$ Monte Carlo
normalized to the integrated luminosity of the data.
This plot is for the combined data set $\roots = 130$-$172$ GeV.
} 
\label{f:x_costh}
\end{figure}

\begin{figure}[b]
\centerline{\epsfig{file=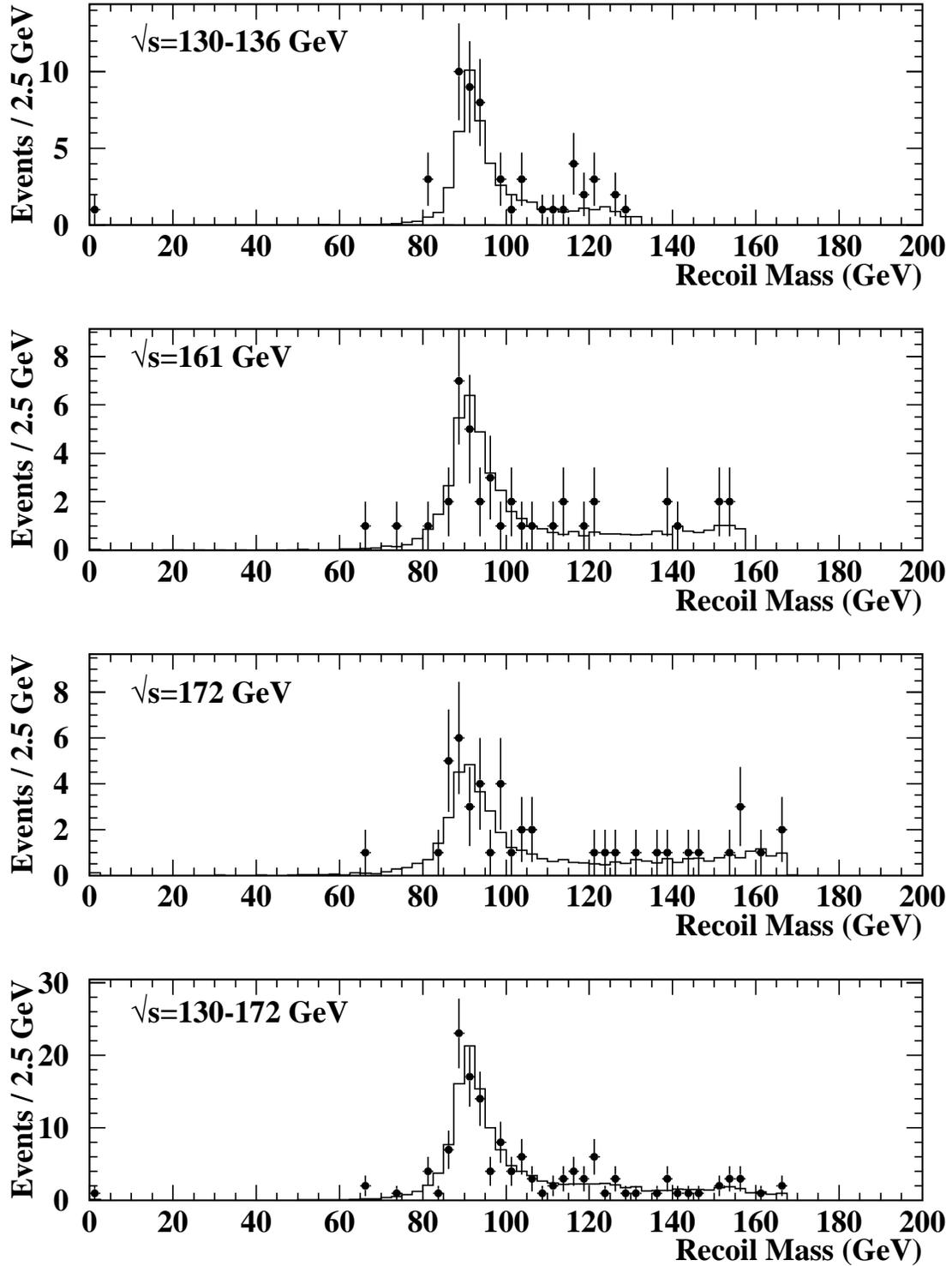,height=22cm,
bbllx=25pt,bblly=20pt,bburx=540pt,bbury=780pt}}
\caption{ The recoil mass distribution for events passing the
single photon selection for the $\roots$ = 130-136 GeV, 
161 GeV, 172 GeV, and combined
$130$ - $172$ GeV data samples. 
The points with error bars are the data and the histograms are the
expectations from the KORALZ $\eetonnggbra$ Monte Carlo
normalized to the integrated luminosity of the data. 
}
\label{f:recm}
\end{figure}

In Figure~\ref{f:x_costh}a, the scaled energy 
of the most energetic photon is plotted against the cosine
of its polar angle for events in the $\roots$=172 GeV sample.
The data are distributed as expected from the
$\eetonnggbra$ Monte Carlo. Similar agreement is seen for
the 130, 136 and 161 GeV data. In Figure~\ref{f:x_costh}b the 
polar angle distribution for the entire $\roots$=130-172 GeV
sample is shown and agrees with
the $\eetonnggbra$ Monte Carlo expectation.
If one calculates the mass recoiling against the photon
(or against the two-photon system) in these events,
one expects a peak in the recoil mass at $\mathrm{M}_{Z}$,
since the $\nunu$ predominantly comes from the decay of a $\PZz$.
One clearly sees this feature in the data as shown in Figure~\ref{f:recm}.
In general there is also
good agreement between data and Monte Carlo in this distribution.
However, at $\sqrt{s}=136$ GeV, there is one event 
with a measured photon energy
of 84~GeV.
Its estimated recoil mass is imaginary as the 
measured energy exceeds the beam energy, and it 
is shown in the distribution as occuring at zero recoil mass.
A careful study of this event shows strong evidence,
besides the measured energy, that it
comes from a cosmic ray, well out-of-time with respect to the
LEP beam crossing. In fact, it is sufficiently out-of-time so
as to miss detection in several OPAL detector elements including
the TOF system. It is left in the data sample because it passes all 
the selection criteria.  It is not, however,  
considered to be a good physics event candidate.

The single-photon selection was designed to allow for the presence
of a second photon in order to accept events from the $\eetonngggbra$
process.
In the data 12 out of the 138 selected events are considered to
be two photon events (i.e., have a second photon with deposited
energy exceeding 300~MeV). This is consistent with the expectation  
of 7.1 events from the KORALZ Monte Carlo. Ten of the 12 events are in 
common with the acoplanar-photons event selection.

\boldmath
\subsubsection{Search for $\eetoXY$, $\XtoYg$ - General case: $\my\ge 0$}
\unboldmath
\label{sec:g1_results_allmy}

Selected events at a given centre-of-mass energy
are classified as consistent with a given value of $\mx$ and $\my$
if the energy of the most energetic photon falls within the region 
kinematically accessible to a photon from the process $\eetoXY$, 
$\XtoYg$. The kinematic consistency criterion includes allowance for the 
energy resolution and incurs an inefficiency of less than 2\% for all
values of $\PX$ and $\PY$ masses while
accepting only those Standard Model $\eetonnggbra$ 
events that are kinematically consistent with 
a given $\PX$ and $\PY$ mass hypothesis.
 
The kinematic region with 
true recoil mass significantly below $M_{\rm Z}$
has only a small background expectation
from Standard Model $\eetonnggbra$,
but it may be populated as a result of energy mis-measurement
in  regions of the detector with poorer energy resolution. 
We reject events as candidates for XY production if the 
most energetic photon is in one of the following 
angular regions, $0.785 < \acosthe < 0.815$ and 
$\acosthe > 0.94$, and the recoil mass is below 75 GeV.
One data event is rejected by this cut
compared with 
1.4 events expected from the 
Standard Model $\eetonnggbra$ process.
We have studied the 
effect of further
cuts to reduce the Standard Model $\eetonnggbra$ background.
We find that a significant improvement in the expected 
sensitivity can be achieved for small
$\my$
by accepting only events with recoil mass significantly below 
$M_{\rm Z}$. 
Events are retained as candidates 
for small $\my$, defined as 
$\my < 14 +0.1\mx$~(GeV), if the measured recoil mass is less 
than 75 GeV. For the complementary, large $\my$  
region, $\my > 14 +0.1\mx$ (GeV),
no other cuts are applied. 
The boundary between the small and large $\my$ region
was chosen so as to optimise
the expected sensitivity\footnote{The optimisation condition
chosen was that the upper limit that one would 
expect to set, on average, in the absence of 
new physics contributions should be minimised.
This definition has the advantage that it 
does not require one to specify the 
cross-section of possible new physics.}  
for the
combined data sample.
For simplicity, the same boundary 
is
applied at each centre-of-mass energy.
 
\begin{figure}[ht]
        \centerline{\epsffile{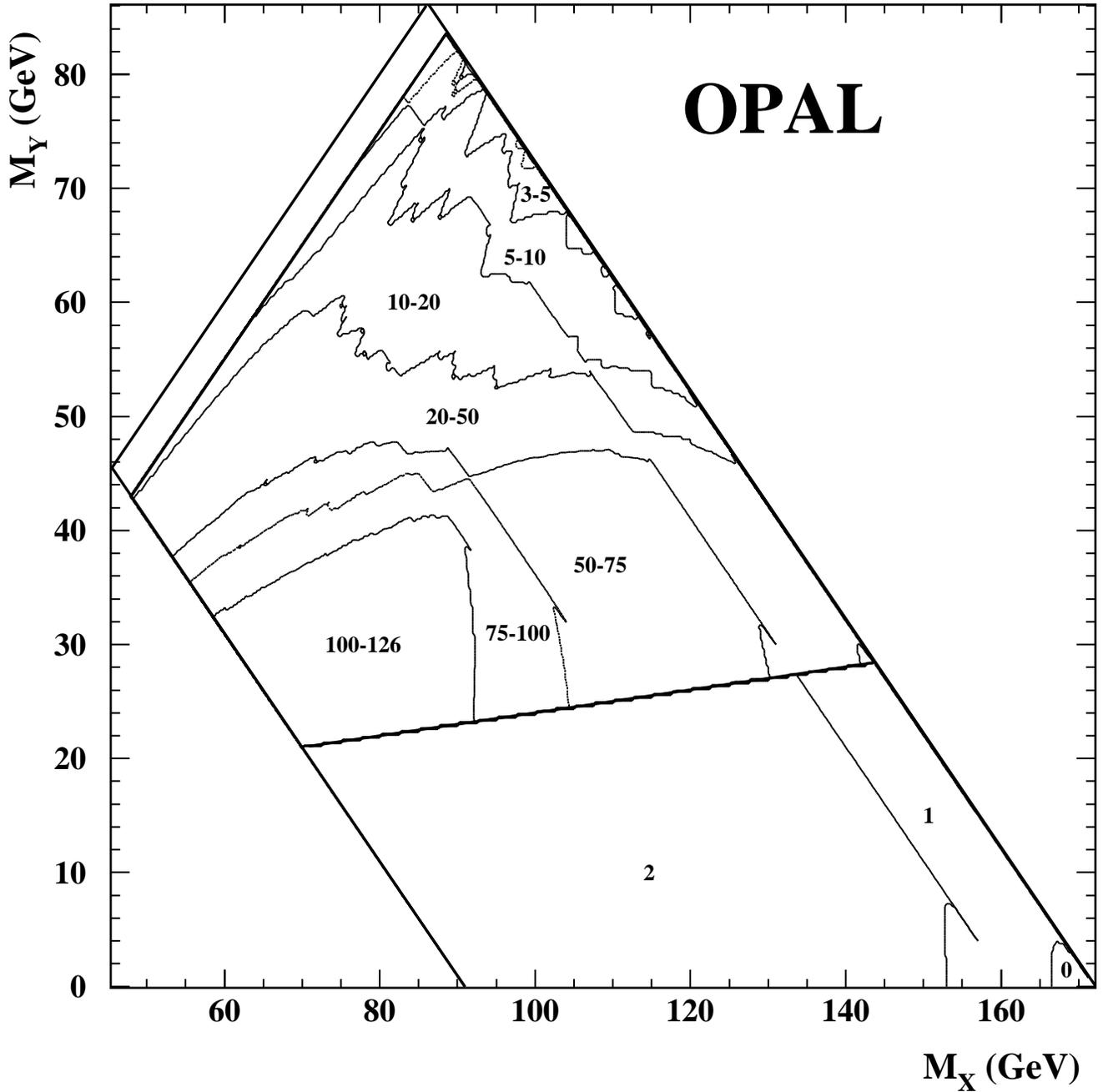}}
        \caption{Number of single photon candidate events 
                in the combined data sample ($\sqrt{s} =$ 130-172 GeV) 
                consistent with each set of mass values ($\mx$, $\my$)
                for the process $\eetoXY$, $\XtoYg$,
                after application of all selection criteria including
                kinematic consistency requirements.
                Lines are drawn around the boundaries defined by 
                $\mx + \my = 172$ GeV, $\mx = \my$, and 
                $\mx + \my = M_{\rm Z}$, and to display the
                boundary  
                between the small and large $\my$ regions.  Regions
                defined by $\mx - \my < 5$ GeV and 
                $\mx + \my < M_{\rm Z}$
                are not considered in limit calculations, and no events
                are displayed in these regions.}
        \label{fig:g1_evsXY_data}
\end{figure}

\begin{figure}[ht]
        \centerline{\epsffile{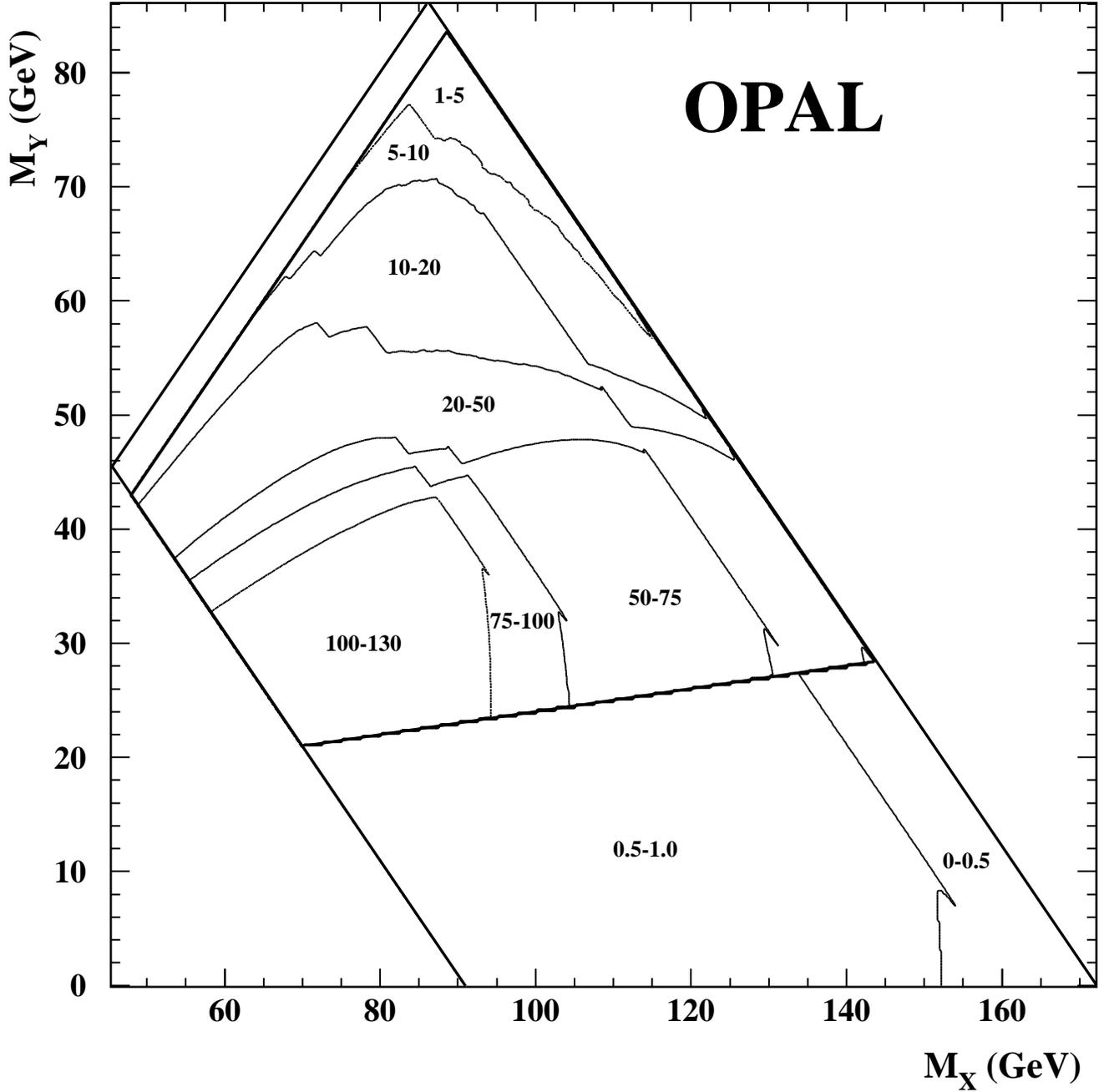}}
        \caption{
                Number of Standard Model $\nu \bar{\nu} \gamma (\gamma)$ events 
                predicted by KORALZ to pass all single photon selection 
                criteria for the combined data sample 
                ($\sqrt{s} =$ 130-172 GeV),
                which are consistent with the process 
                $\eetoXY$, $\XtoYg$ at each set of
                mass values ($\mx$, $\my$).
                This figure gives the expected Standard Model 
                contribution to Figure~\ref{fig:g1_evsXY_data}.
                Boundaries and delineated regions are the same as in 
                Figure~\ref{fig:g1_evsXY_data}.}
        \label{fig:g1_evsXY_mc}
\end{figure}

\begin{figure}[ht]
        \centerline{\epsffile{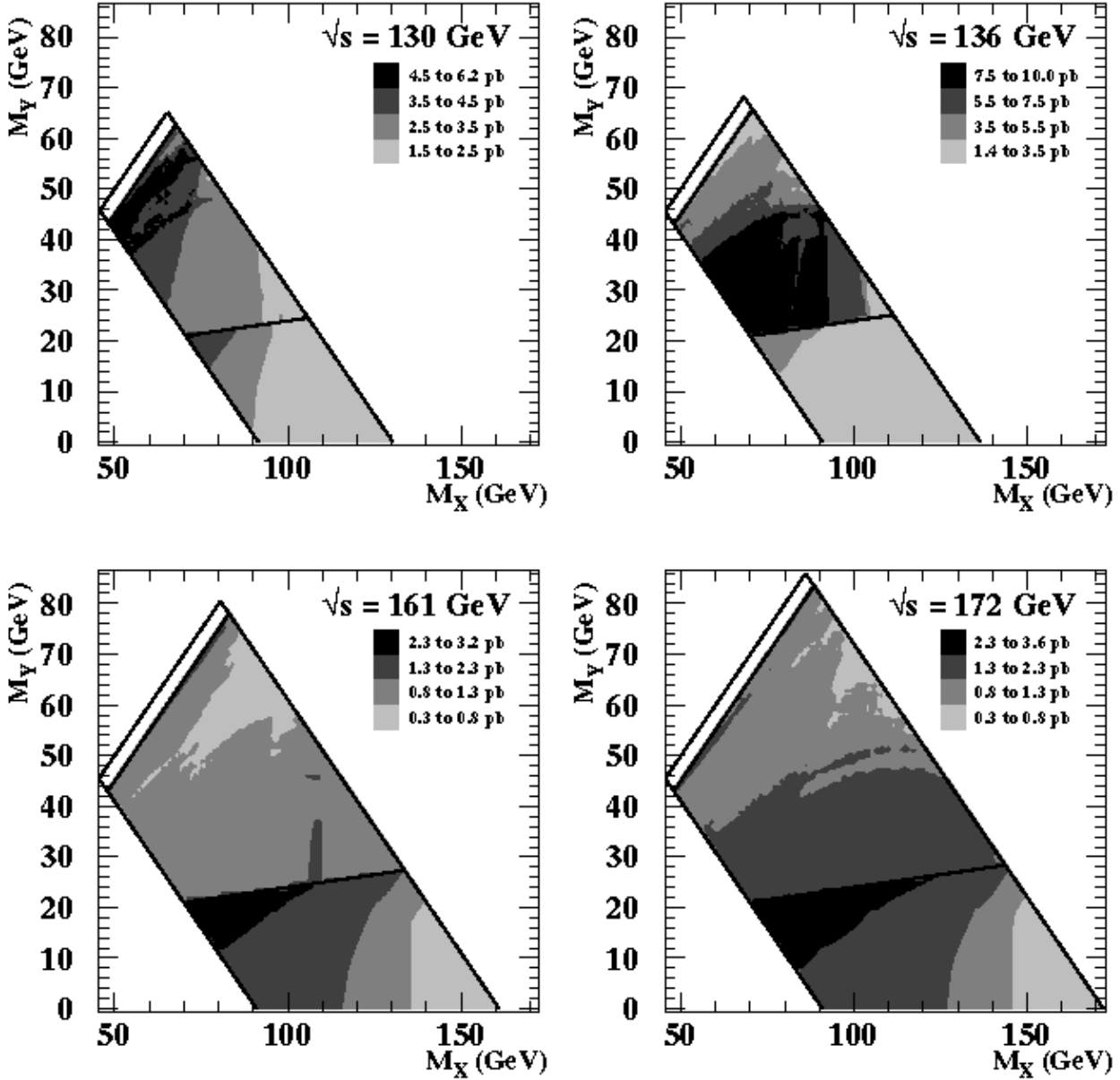}}
        \caption{95\% CL upper limits on 
                $\sigbrXY$
                as a function of $\mx$  and $\my$
                at each centre-of-mass energy.
                Lines are drawn to indicate the kinematically allowed
                boundaries defined by ${\mx + \my} = \sqrt{s}$.
                Other boundaries and delineated regions are as defined for 
                Figure~\ref{fig:g1_evsXY_data}.}
        \label{fig:g1_XYlim_each}
\end{figure}

\begin{figure}[ht]
        \centerline{\epsffile{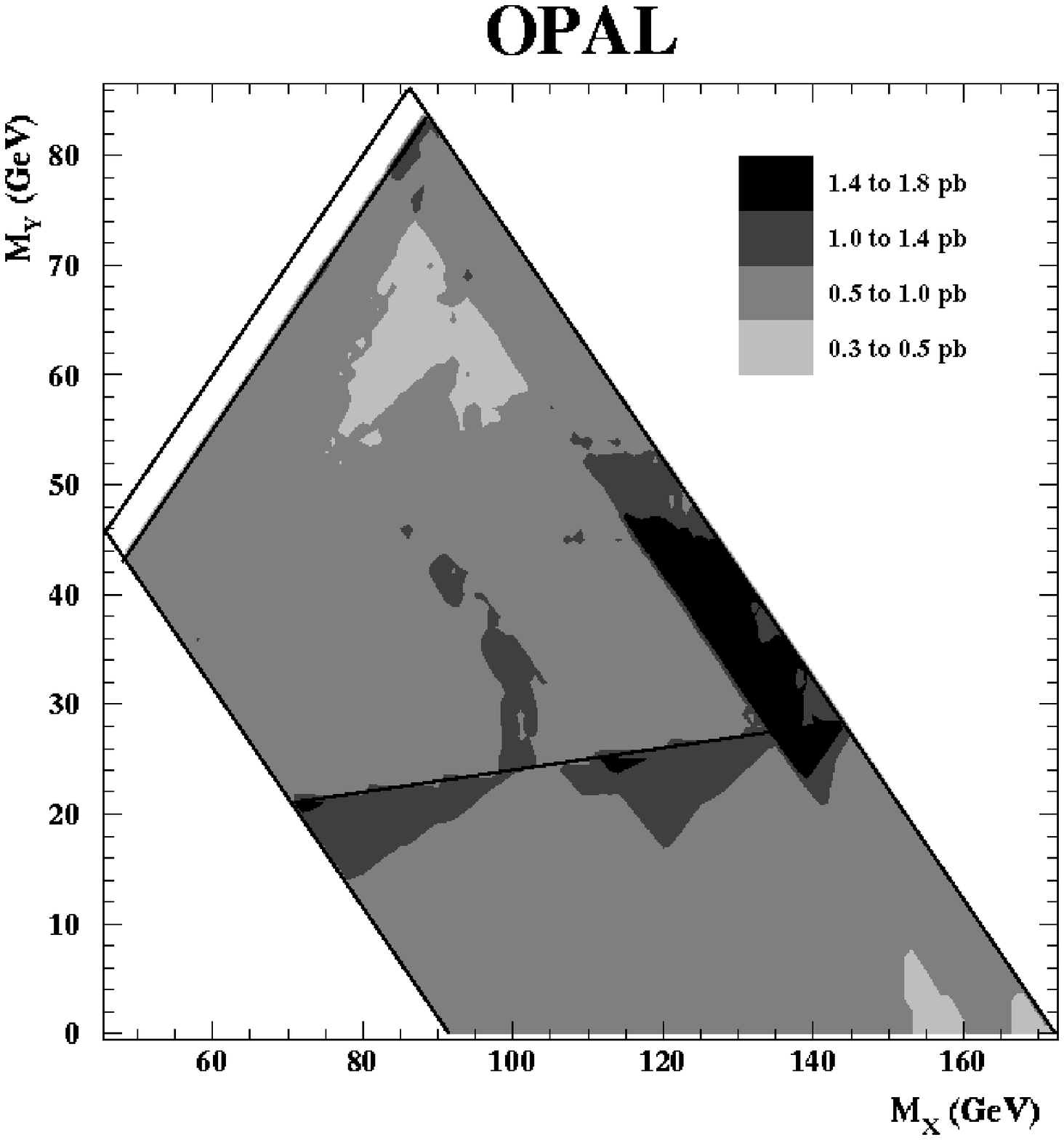}}
        \caption{The 95\% CL upper limit on $\sigbrXY$ 
                at $\roots = 172$ GeV
                as a function of $\mx$  and $\my$, 
                obtained from the combined data sample assuming a
                cross-section scaling of $\betax/s$.
                The boundaries and delineated regions are as defined for 
                Figure~\ref{fig:g1_evsXY_data}.}
        \label{fig:g1_XYlim_all}
\end{figure}

The selection efficiencies at each 
generated grid point for the SUSYGEN 
Monte Carlo events are shown for $\roots = 130$ GeV in Table~\ref{tab:g1_eff130} 
and for $\roots = 172$ GeV in Table~\ref{tab:g1_eff172}.  
The efficiencies at intermediate centre-of-mass energies lie between
those shown for these two centre-of-mass energies.
These values include the efficiency of the kinematic consistency 
selection criteria which is higher than 98\% 
at each generated mass point. 
The number of selected events consistent with 
each $(\mx,\my)$ value is shown in Figure~\ref{fig:g1_evsXY_data}
and can be compared with the number expected from Standard 
Model $\eetonnggbra$ 
events as shown in Figure~\ref{fig:g1_evsXY_mc}.
The background  event described earlier does not survive the kinematic
consistency cuts for any kinematically accessible 
point in the $(\mx,\my)$ plane.
In general there is good agreement,
and we proceed to set upper limits at 95\% CL 
on the cross-section times branching ratio $\sigbrXY$.
These upper limits are first calculated separately at each 
centre-of-mass energy, as shown in Figure~\ref{fig:g1_XYlim_each}.
A combined upper limit on the cross-section times branching ratio at $\sqrt{s}$
of 172 GeV is calculated combining the information from 
each centre-of-mass energy. The combination is performed 
assuming a cross-section centre-of-mass energy dependence
of $\betax/s$, where $\betax$ is the speed of particle $\PX$. 
This combined limit is shown in Figure~\ref{fig:g1_XYlim_all}.
The upper limits are calculated taking into account the 
expected number of Standard Model $\eetonnggbra$ background events 
estimated from KORALZ using the method described in \cite{rf:PDG96}.
The estimated non-physics background is
intentionally not taken into account in the limit calculation.
The resulting combined upper limits range from 0.31 pb to 1.8 pb.

Systematic errors are due primarily to limited Monte Carlo statistics at the generated
($\mx,\my$) points and the uncertainty on the efficiency parametrization across the 
($\mx,\my$) plane. The combined relative uncertainty on the efficiency is 4\%.
The effect of this uncertainty on the upper limits is 
calculated according to \cite{rf:systerr} and is
found to be negligible.
\boldmath
\subsubsection{Search for $\eetoXY$, $\XtoYg$ - Special case: $\myzero$}
\unboldmath
\label{sec:g1_results_my0}

\begin{figure}[ht]
        \centerline{\epsffile{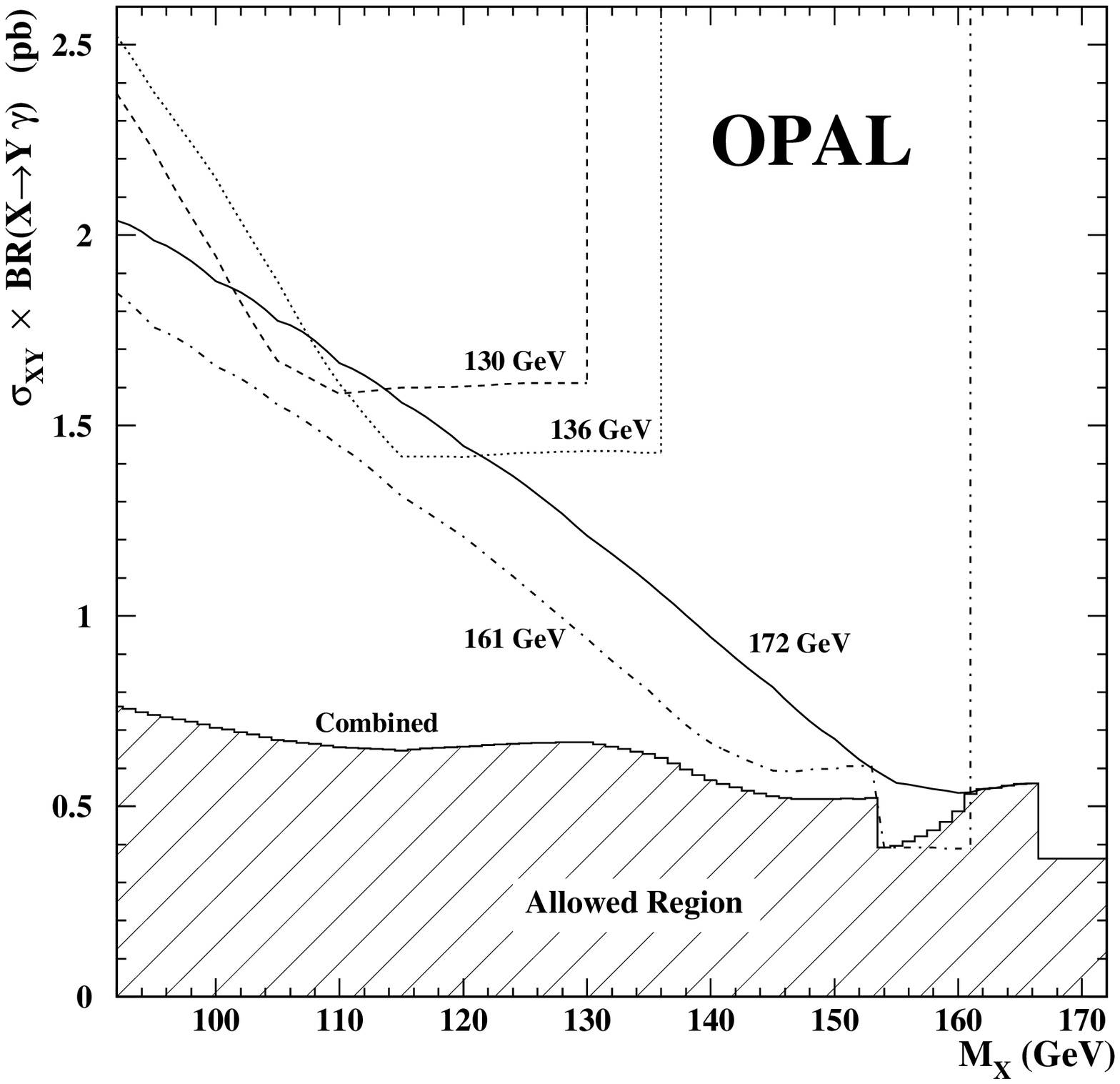}}
        \caption{The 95\% CL upper limits on $\sigbrXY$ 
                as a function of $\mx$, assuming $\myzero$, for each 
                centre-of-mass energy. The upper limit
                on $\sigbrXY$, evaluated at $\sqrt{s}=172$ GeV, 
                obtained from the combined data sample
                is also shown. The combination was performed assuming a 
                cross-section scaling of $\betax/s$. The allowed region is shaded.}
        \label{fig:g1_XYlim_massless}
\end{figure}

The above results
include the case of $\myzero$ that
lies within the high mass-difference region in which the 
expected number of events is small. 
For example, for $\mx=100$~GeV, two events are observed
compared with an expected contribution from $\nnggbra$ of $0.8\pm{0.1}$ events.
For $\mx=170$~GeV, no events are observed; the background expectation is 
$0.10\pm{0.03}$ events.
In this region, the requirement that the recoil-mass be less than 
75~GeV eliminates all sources of physics background except a small  residual
contribution 
from $\eetogg$($\gamma$); the expected contribution is about 0.01 events and is 
neglected. The upper limits for the $\myzero$ case, as a function of $\mx$, are shown in 
Figure~\ref{fig:g1_XYlim_massless}. The resulting combined upper limits on 
$\sigbrXY$ range from 0.36 pb to 0.76 pb. 
Interpretation of these results for the production of excited neutrinos is described
in a separate paper~\cite{ref:opal_exlp_172}.

\boldmath
\subsection {Acoplanar-Photons $\gamma\gamma(\gamma)+{E_T\hspace*{-1.0em}/\quad}$ }
\unboldmath
\label{sec:g2_results}

After applying the acoplanar-photons selection criteria to the combined data sample,
a total of 11 events are observed.  
The predictions for the total number of Standard Model $\eetonngggbra$ events
is $6.3\pm{0.2}$, based on the KORALZ generator.
Non-physics background as well as contributions from other Standard Model 
processes are negligible. The breakdown of the observed 
and expected number of events for the different centre-of-mass energies is given in 
Table~\ref{tab:g2_nevt}. Within the kinematic acceptance, the selection 
efficiency\footnote{Before accounting for efficiency losses 
due to detector occupancy and 
detector status requirements.}
for $\eetonngggbra$ events is rather constant as a function
of centre-of-mass energy, varying from about (67-70)\% for the KORALZ generator and 
from (67-72)\% for NNGG03 and NUNUGPV. The mean efficiencies are ($68.6\pm{1.5}$)\%
for KORALZ and ($70.4\pm{0.5}$)\% for NNGG03 and NUNUGPV. As the 
cross-section measurements are 
statistics limited, an efficiency of ($69\pm{3})\%$ is used independent of $\roots$. 
Additional systematic errors arise due to the energy scale for low-energy photons (5\%),
and from uncertainty on the luminosity measurement ($<1\%$).
The measured cross-sections for the process $\eetonngggbra$, within the kinematic 
acceptance defined by the energy and polar angle selection criteria described earlier,
are included in  Table~\ref{tab:g2_nevt} as are the cross-section predictions 
from the KORALZ generator.

\begin{figure}[b]
\centerline{\epsfig{file=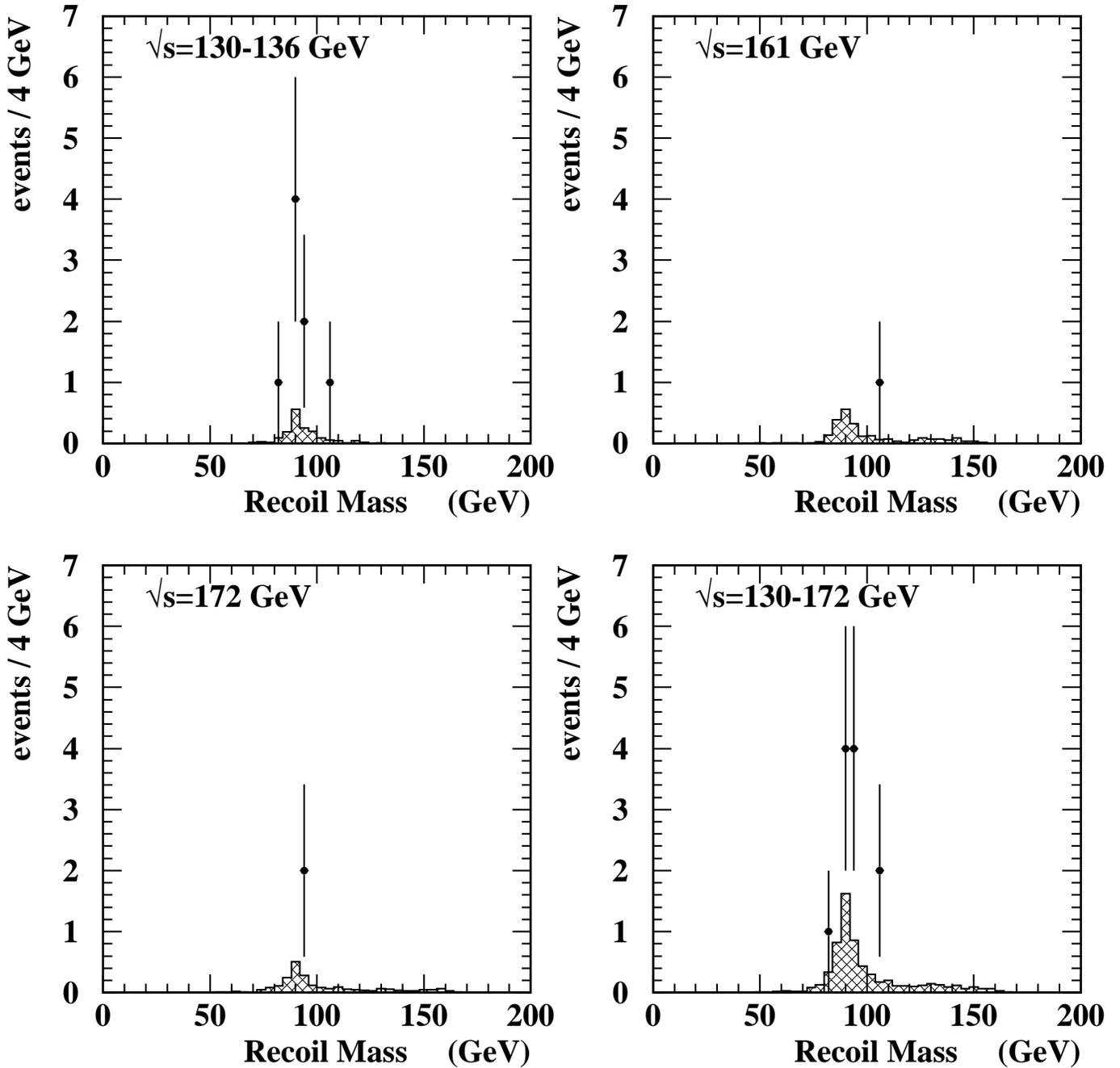,width=11cm,
bbllx=125pt,bblly=150pt,bburx=430pt,bbury=680pt}}
\caption{Recoil-mass distributions for the selected acoplanar-photons events
for each centre-of-mass region and for the combined data sample.
The data points with error bars represent the selected OPAL data events.
In each case the shaded histogram shows the expected contribution from 
$\eetonngggbra$ events, normalized to the 
total integrated luminosity. The KORALZ generator was used for these distributions.
}
\label{fig:g2_mrec}
\end{figure}

\begin{figure}[b]
\centerline{\epsfig{file=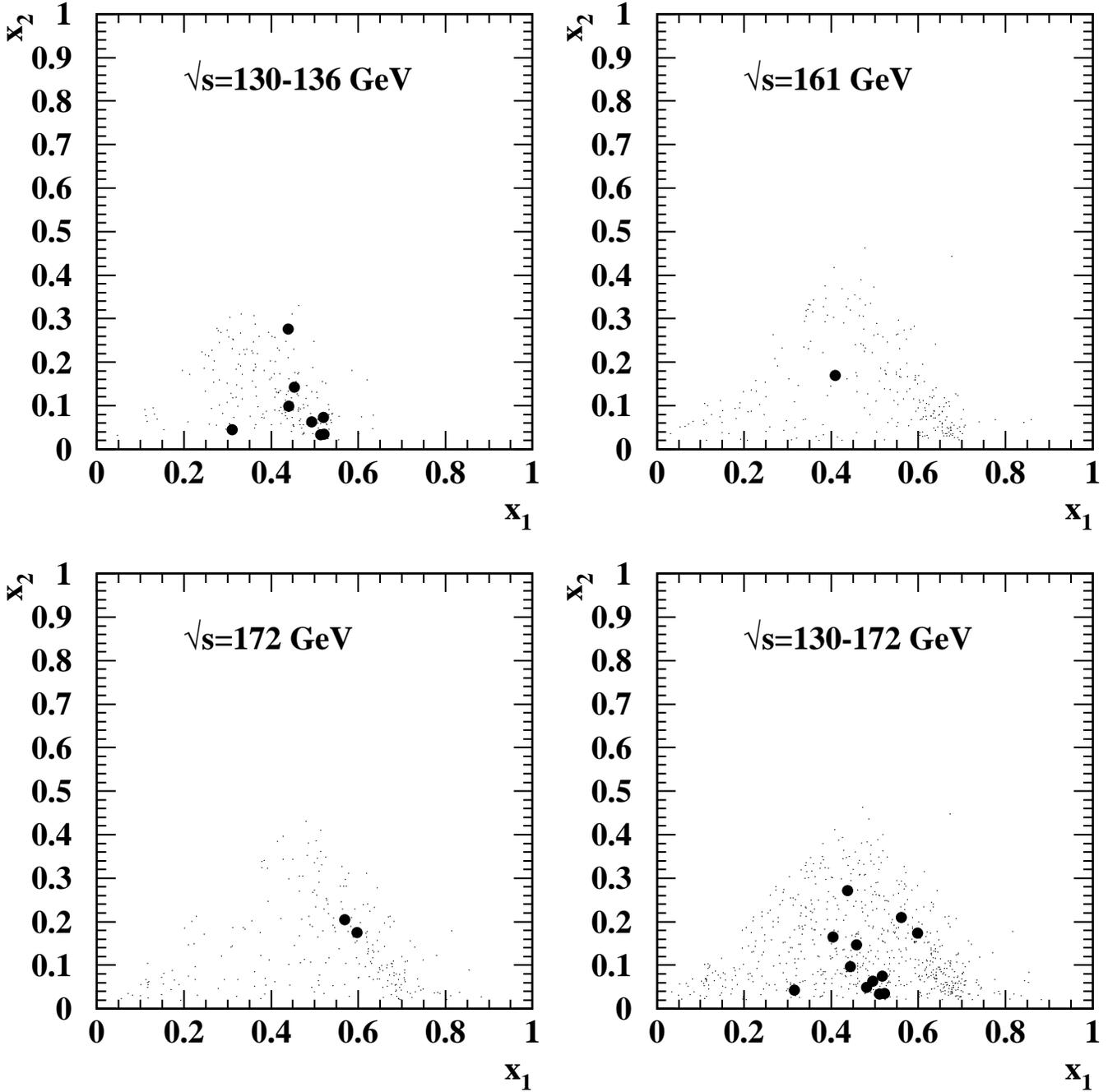,width=11cm,
bbllx=125pt,bblly=150pt,bburx=430pt,bbury=680pt}}
\caption{$x_{2}$ vs  $x_{1}$ for  
the selected acoplanar-photons events
for each centre-of-mass region and for the combined data sample.
The large points represent the selected OPAL data events.
The smaller dots show the expected distribution for events from the process
$\eetonngggbra$ events, from KORALZ.
The normalization of the Monte Carlo distributions is arbitrary. However,
for the combined plot the relative normalizations
of the various data samples is maintained.
}
\label{fig:g2_x1x2}
\end{figure}

\begin{figure}[b]
\centerline{\epsfig{file=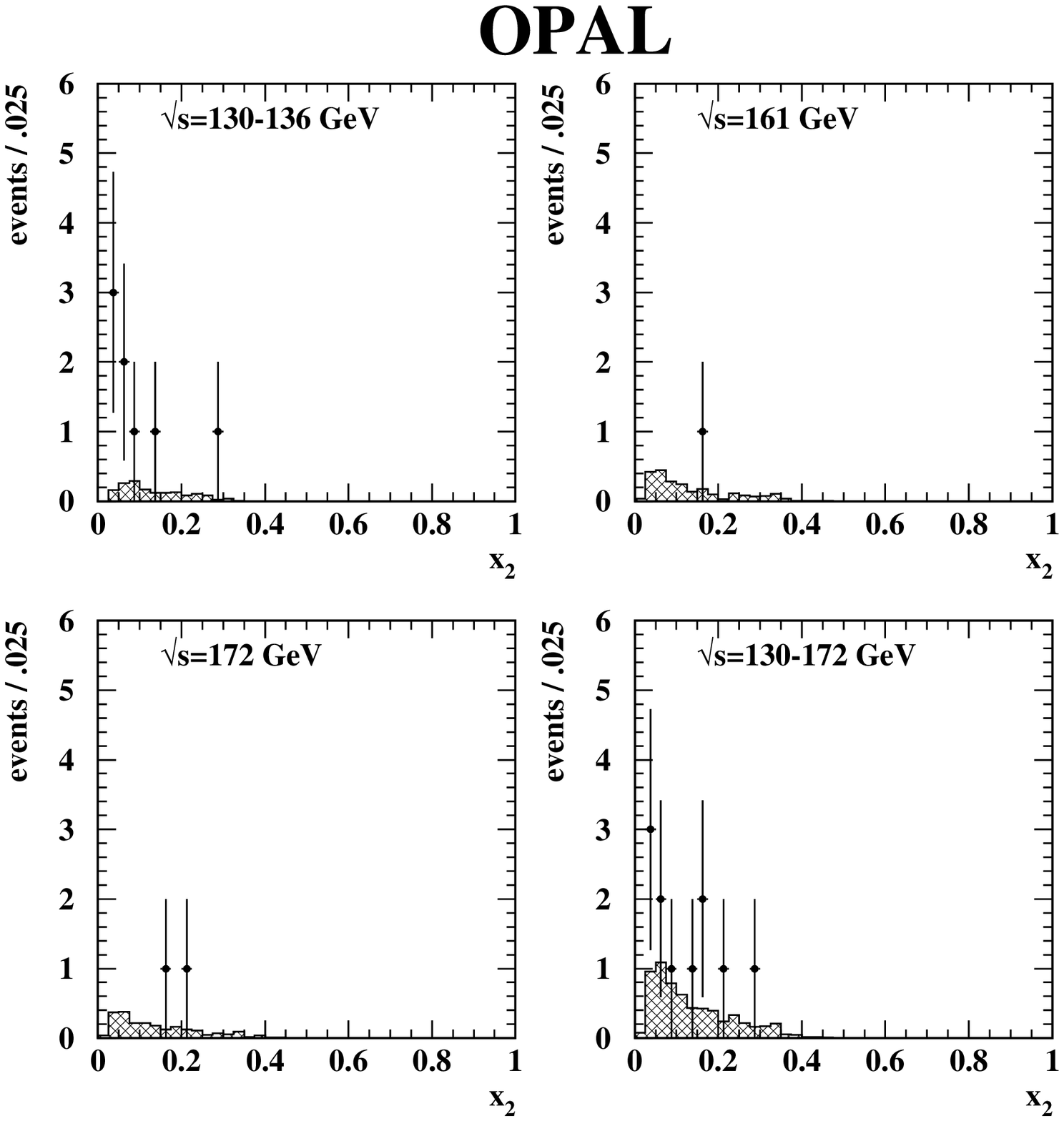,width=11cm,
bbllx=125pt,bblly=150pt,bburx=430pt,bbury=680pt}}
\caption{Distributions of scaled energy of the lower energy photon 
for the selected acoplanar-photons events
for each centre-of-mass region and for the combined data sample.
The data points with error bars represent the selected OPAL data events.
In each case the shaded histogram shows the expected contribution from 
$\eetonngggbra$ events, normalized to the 
total integrated luminosity. The KORALZ generator was used for these distributions.
}
\label{fig:g2_x2}
\end{figure}

\begin{figure}[b]
\centerline{\epsfig{file=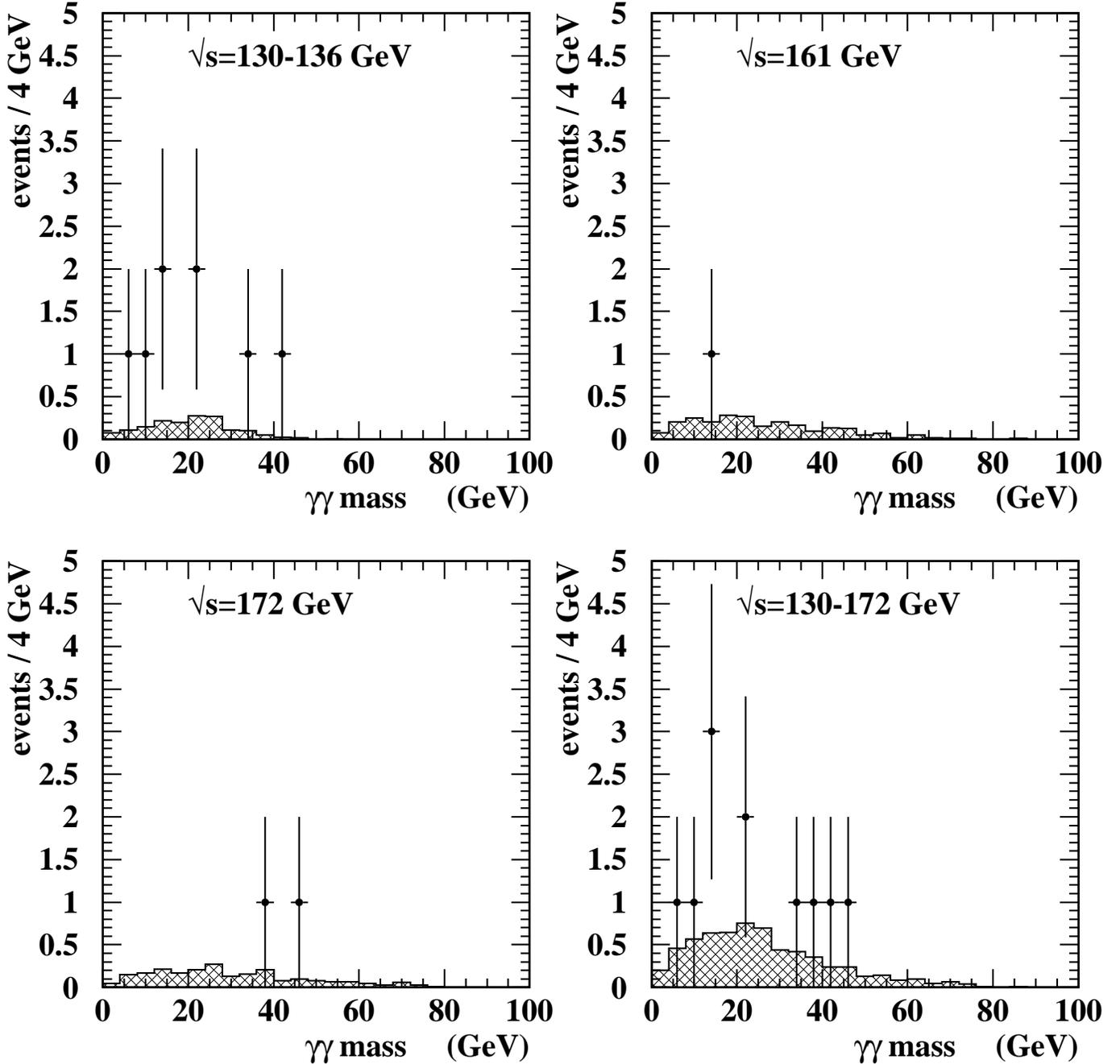,width=11cm,
bbllx=125pt,bblly=150pt,bburx=430pt,bbury=680pt}}
\caption{Distributions of the $\gamma\gamma$ invariant mass of the accepted 
acoplanar-photon pairs for each centre-of-mass region. The points with
error bars represent the OPAL data while the shaded histograms show the 
predicted distributions for events from 
$\eetonngggbra$ events, from KORALZ,
normalized to the corresponding integrated luminosity.
}
\label{fig:g2_ymgg}
\end{figure}

There were no selected events having $N_{\gamma}>2$, compared with an 
expectation from KORALZ of $0.34\pm{0.06}$ events.
The kinematic properties of the selected events are summarized in 
Table~\ref{tab:g2_kine} and displayed in Figures~\ref{fig:g2_mrec}-\ref{fig:g2_ymgg} 
where they are  compared with the predicted distributions for $\eetonngggbra$, 
obtained using the KORALZ generator, normalized to the corresponding 
integrated luminosity. Figure~\ref{fig:g2_mrec} shows the recoil-mass distributions
of the selected acoplanar-photon pairs. These are peaked near the mass of the $\PZz$
as expected for contributions from $\eetonngggbra$. 
The resolution of the recoil mass is typically about 2-3 GeV
at each centre-of-mass energy for $M_{\rm recoil}\approx M_{\rm Z}$. 
Figure~\ref{fig:g2_x1x2} shows 
the $x_2$ vs. $x_1$ distributions for each centre-of-mass energy
and for the combined data sample.
The projections of the scaled energy of the least-energetic photon are given in 
in Figure~\ref{fig:g2_x2}. 
Figure~\ref{fig:g2_ymgg} shows the distributions of the invariant mass, 
$M_{\gamma\gamma}$,
for the selected acoplanar-photon pairs at each centre-of-mass energy. 
The mass resolution is typically 0.6-1.4 GeV. A search
for $\rm H^0\rightarrow\gamma\gamma$ has been recently published by OPAL\cite{OPAL_Hgg}.
  
For the data at $\roots = 161$, $172$ GeV, the measured distributions 
agree with the $\nngggbra$ expectation. For the 
$\roots=130,136$ GeV data there is an apparent excess of events. However, 
with the exception of the $x_2$ distribution in Figure~\ref{fig:g2_x2}, for
which the measured distribution peaks more strongly than expected at low values,  
the distributions appear consistent in shape with the expectation from 
$\eetonngggbra$. This excess was also  remarked on in a 
previous publication\cite{rf:OPALSP130}.


\boldmath
\subsubsection{Search for $\eetoXX$, $\XtoYg$ - General case: $\my\ge 0$}
\unboldmath
\label{sec:g2_results_allmy}

Selected events are classified as consistent with a given value of $\mx$ and $\my$
if each of the two selected photons falls within the region kinematically accessible 
to photons from the process $\eetoXX$, $\XtoYg$. 
As before, this includes allowance for resolution effects. 
Monte Carlo events were generated at each centre-of-mass 
energy. The selection efficiencies at each generated grid point for the 
$\roots = 172$ GeV SUSYGEN Monte Carlo events are shown in Table~\ref{tab:g2_eff172}. 
These values include 
the efficiency of the kinematic consistency selection criteria 
which is higher than 95\% at each generated point in the $(\mx,\my)$ plane.
Similar efficiencies are obtained at the other centre-of-mass energies.

\begin{figure}[b]
\centerline{\epsfig{file=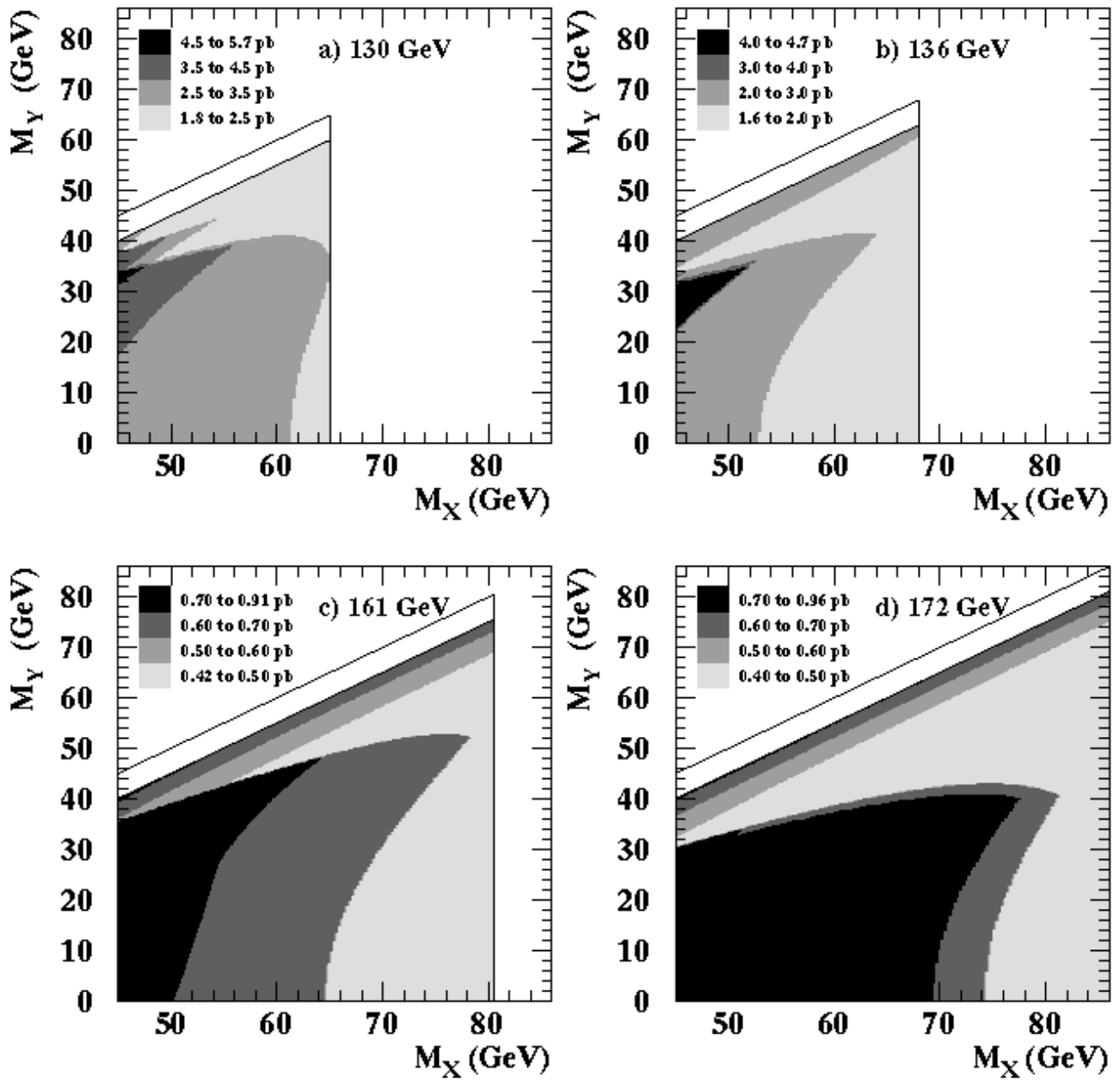,width=11cm,
bbllx=125pt,bblly=150pt,bburx=430pt,bbury=680pt}}
\caption{For each centre-of-mass energy,
the shaded areas show 95\% CL exclusion regions for 
$\sigbrXX$.
No limit is set for mass-difference values 
$\mx-\my < 5$ GeV, defined by the lower line above 
the shaded regions. The upper line is for  $\mx=\my$.
}
\label{fig:mxmy_each}
\end{figure}

\begin{figure}[b]
\centerline{\epsfig{file=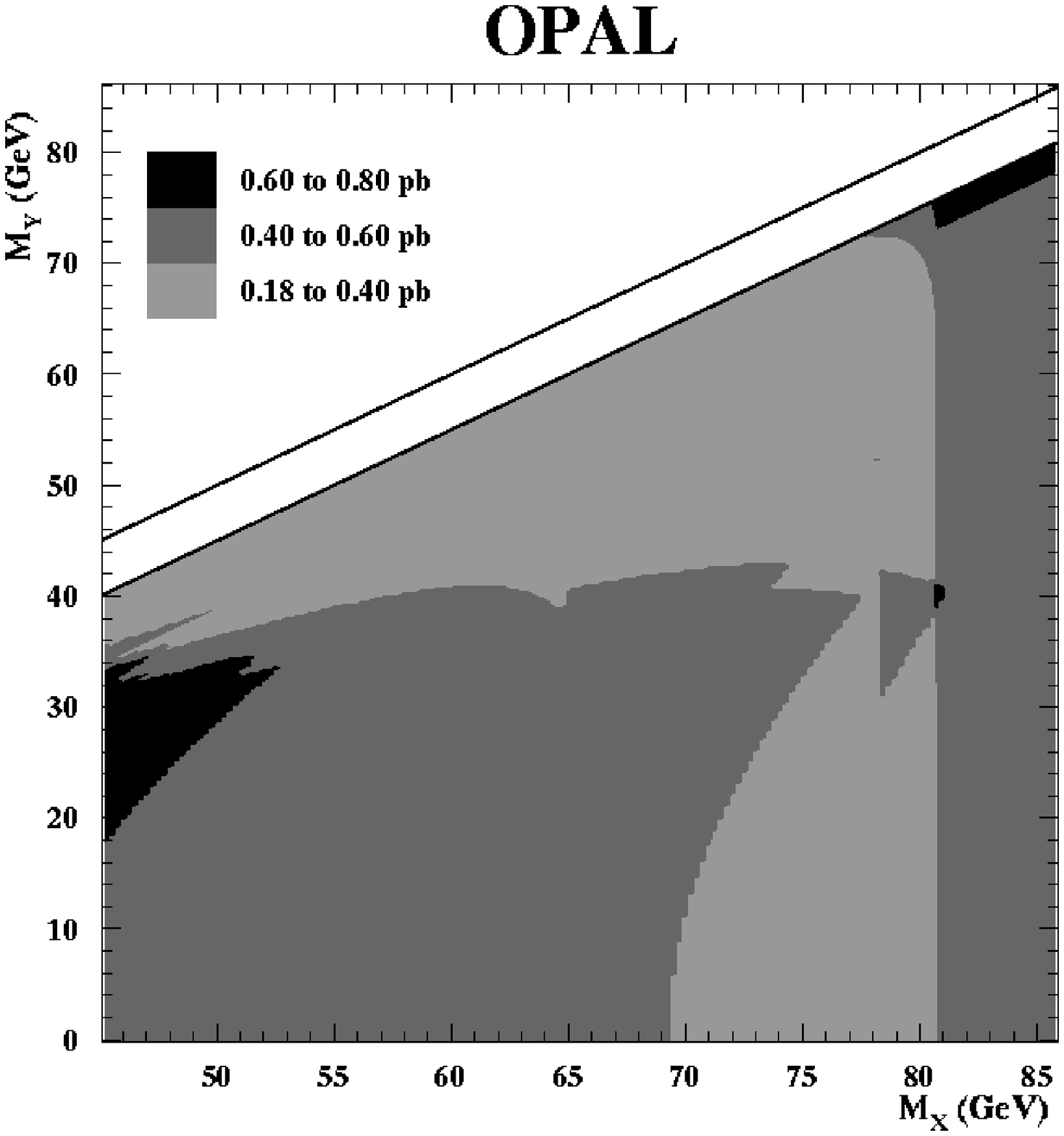,width=11cm,
bbllx=125pt,bblly=150pt,bburx=430pt,bbury=680pt}}
\caption{
The shaded areas show 95\% CL exclusion regions
for $\sigbrXX$  at $\roots = 172$ GeV, obtained from
the combined data sample assuming a cross-section scaling
of $\betax/s$. 
No limit is set for mass-difference values 
$\mx-\my < 5$, defined by the lower line above 
the shaded regions. The upper line is for $\mx=\my$.
}
\label{fig:mxmy_all}
\end{figure}

Figures~\ref{fig:mxmy_each} (a)-(d) show the 95\% CL $\sigbrXX$
exclusion plots obtained at $\roots= 130$, $136$, 161 and 172 GeV respectively.
Because of the current uncertainties on the modelling of the Standard Model background,
as discussed earlier, these limits and the limits presented below for this topology,
have been calculated without taking into
account the background estimate.
Events from $\eetonngggbra$ are typically
characterized by a high-energy photon from the radiative return to the $\rm Z^0$ and
a second lower energy photon. The kinematic consistency requirements, however,
require that the two photons have energies within the same (kinematically accessible)
region. For this reason, two of the selected events are inconsistent
with any $(\mx,\my)$ point for $\mx\ge 45$ GeV and 
$\mx-\my\ge 5$ GeV.
Figure~\ref{fig:mxmy_all} shows the 95\% CL $\sigbrXX$ exclusion plot
obtained from the combined data sample assuming that 
$\sigXX$ scales with centre-of-mass energy as $\betax/s$. For the combined plot, 
the maximum value of the limit is 0.80 pb. 
The minimum value is 0.18 pb. 

Systematic errors are due primarily to limited Monte Carlo statistics at the generated
($\mx,\my$) points and the uncertainty on the efficiency parametrization across the 
($\mx,\my$) plane. The combined relative uncertainty on the efficiency varies from about 
(3-6)\% across the plane. 
All systematic uncertainties are accounted for in the manner advocated in 
reference\cite{rf:systerr}.
This also applies to the limits for the $\myzero$ case, 
presented in the next section.

\boldmath
\subsubsection{Search for $\eetoXX$, $\XtoYg$ - Special case: $\myzero$}
\unboldmath
\label{sec:g2_results_my0}

\begin{figure}[b]
\vskip -1.0cmcp 
\centerline{\epsfig{file=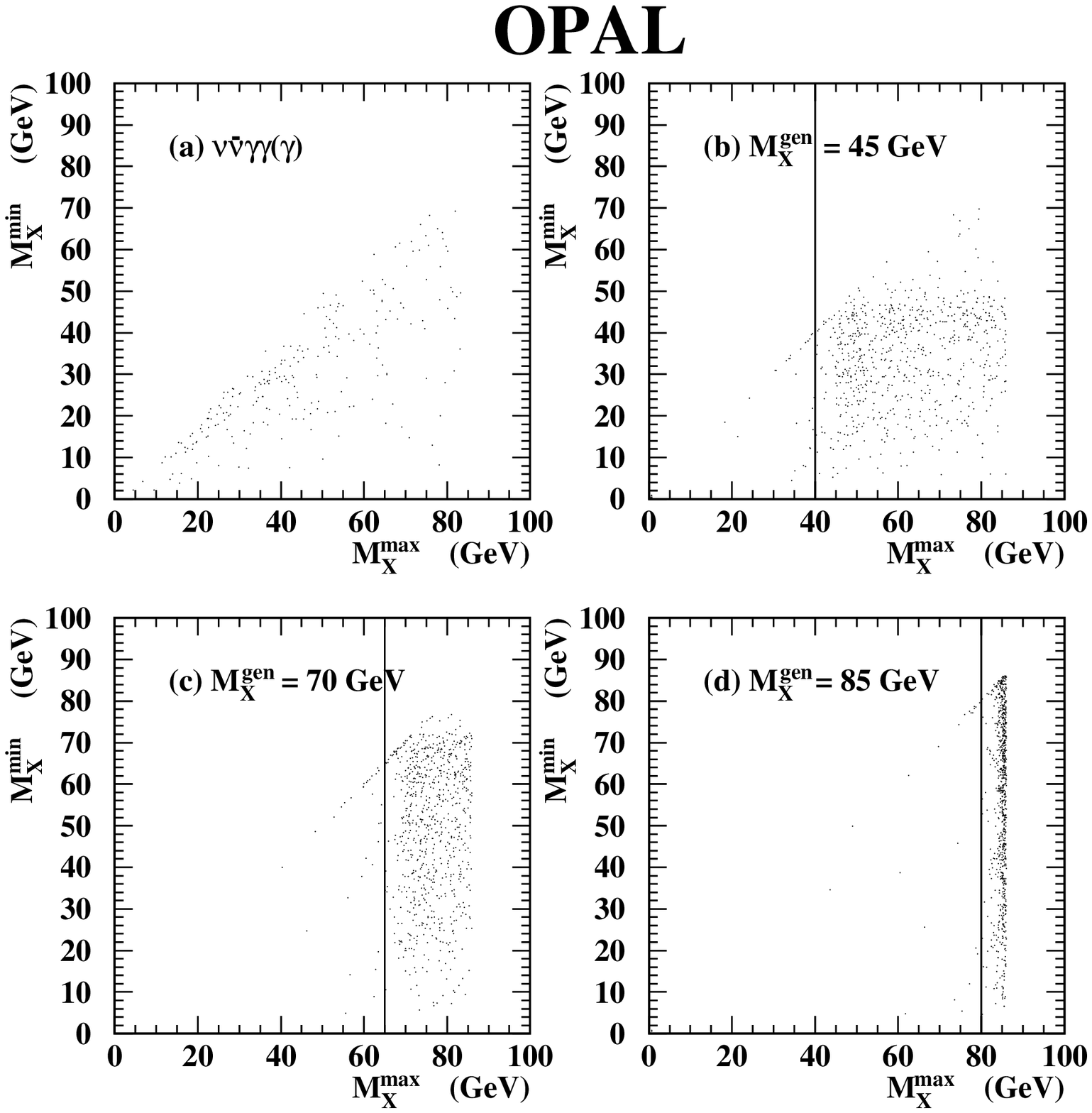,width=11cm,
bbllx=125pt,bblly=100pt,bburx=430pt,bbury=680pt}}
\caption{$\mxmin$ vs. $\mxmax$ for 
(a) $\nngggbra$ Monte Carlo (KORALZ) and (b)-(d)
$\eetoXX$, $\XtoYg$, $\myzero$ Monte Carlo
(SUSYGEN) for various $\mx$. The vertical lines represent the
chosen cut values discussed in the text. These are for Monte Carlo
generated at a centre-of-mass energy of 172 GeV.}
\label{fig:ln172mc}
\end{figure}

For the special case of $\myzero$ the kinematic consistency cuts applied differ
from those used for the general case.
One can calculate\cite{gravitinos2} the maximum and minimum masses,
$\mxmax$ and  $\mxmin$, which are consistent with the kinematic 
properties of the two photons, assuming a massless $\PY$.
As this argument is based only on kinematics it applies generally  
to the case where the acoplanar photon pair originates from pair production of heavy 
neutral particles which decay radiatively to massless invisible final states;
$\eetoXX$ followed by $\XtoYg$, $\myzero$. These maximum and minimum mass values can 
provide further suppression of the $\nngggbra$ background while retaining high 
efficiency for the  signal hypothesis. The background suppression achieved with 
kinematic consistency requirements based on this procedure is much better than that
obtained for the general case since, in this case, the full event kinematics are
used. Figure~\ref{fig:ln172mc} shows 
$\mxmin$ vs. $\mxmax$ for events passing the selection 
criteria described in section~\ref{sec:g2_selection} for, (a)
$\nngggbra$ Monte Carlo and (b-d) $\eetoXX$, $\XtoYg$, $\myzero$ 
signal Monte Carlo, at $\roots=172~\GeV$, for three values of $\mx$. 
The signal Monte Carlo distributions are 
dominantly populated at maximum mass values greater than or equal to the 
generated mass of $\PX$ (e.g. $\nu^*$ or $\lsp$). Resolution effects shift some
entries to lower masses. 
Requiring that the maximum kinematically allowed mass be greater than
$\mx-5$ GeV retains more than 95\% relative efficiency for signal at all values 
of $\mx$ while suppressing  41\% to 96\% of the remaining
$\nngggbra$ events. 
Similar efficiencies are obtained at each of the other centre-of-mass energies.

\begin{figure}[b]
\centerline{\epsfig{file=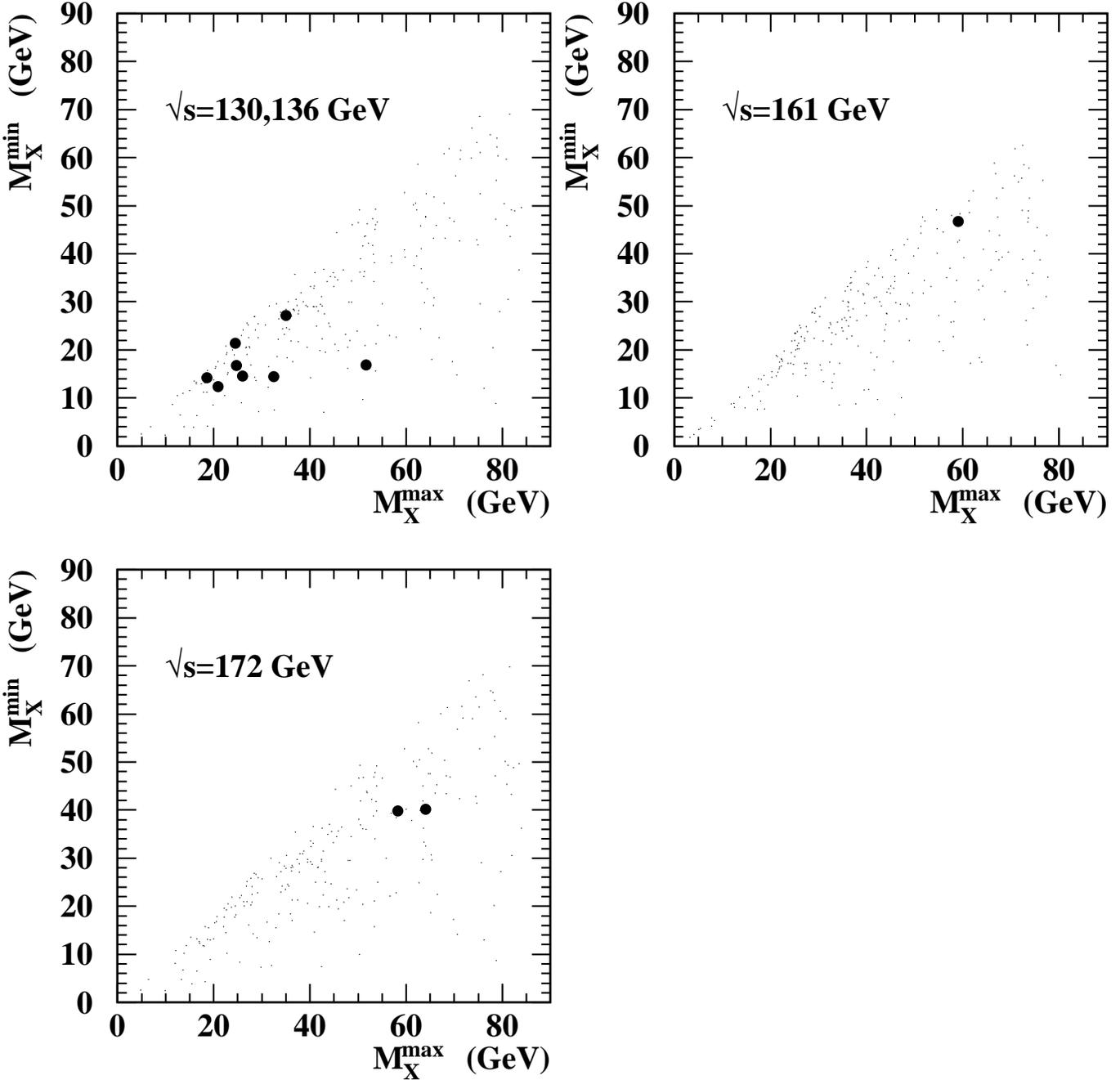,width=11cm,
bbllx=125pt,bblly=150pt,bburx=430pt,bbury=680pt}}
\caption{
$\mxmin$ vs. $\mxmax$ for 
the accepted acoplanar-photons events for each centre-of-mass energy region.
Overlaid are the expected distributions for contributions from the 
Standard Model process 
$\eetonngggbra$, from  
the KORALZ generator.
}
\label{fig:lndata}
\end{figure}

The kinematic properties of the selected events, shown 
in Table~\ref{tab:g2_kine}, include the values of $\mxmax$. 
The data distributions in $\mxmin$ vs. 
$\mxmax$ are shown for each of the centre-of-mass energies in 
Figure~\ref{fig:lndata}. In each case the distribution for 
$\nngggbra$ events is also shown.
For illustrative purposes, the efficiencies calculated from Monte Carlo
events generated at 172 GeV are shown in Table~\ref{tab:g2_eff172_my0} before
and after application of the cut on $\mxmax$. Also shown is
the $\nngggbra$ rejection efficiency obtained with the 
$\mxmax$ cut. Signal efficiencies at other centre-of-mass energies are similar. 

\begin{figure}[b]
\centerline{\epsfig{file=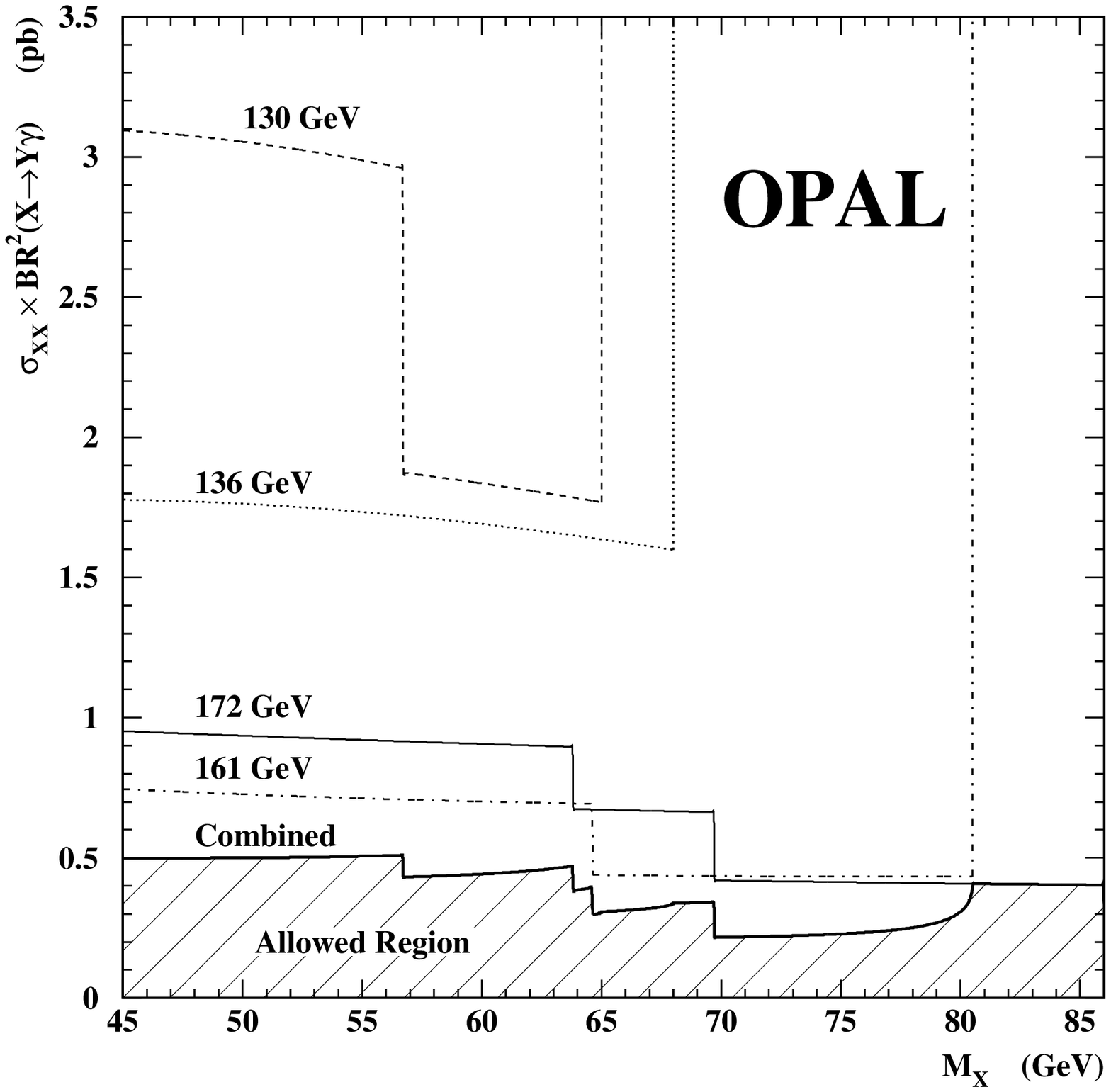,width=11cm,
bbllx=125pt,bblly=150pt,bburx=430pt,bbury=680pt}}
\caption{The 95\% confidence level upper limit on $\sigbrXX$
for the case $\myzero$, as a function of $\mx$, for each value of $\roots$. 
Also shown is the limit, evaluated at $\roots = 172$ GeV, obtained from the 
combined data sample assuming a cross-section scaling of $\betax/s$.
}
\label{fig:limits_each}
\end{figure}

Based on the efficiencies and the 
number of events observed at each centre-of-mass energy, 
we calculate 95\% CL upper 
limits on $\sigbrXX$ (for $\myzero$) 
as a function of $\mx$. 
These are shown in Figure~\ref{fig:limits_each}.
Also shown is the 95\% CL upper limit obtained from the combined data sample,
assuming a centre-of-mass energy dependence of 
the cross-section of $\betax/s$. This combined limit is 0.5 pb or less 
for values of $\mx$ from 45 GeV up to the kinematic limit.

\begin{figure}[b]
\centerline{\epsfig{file=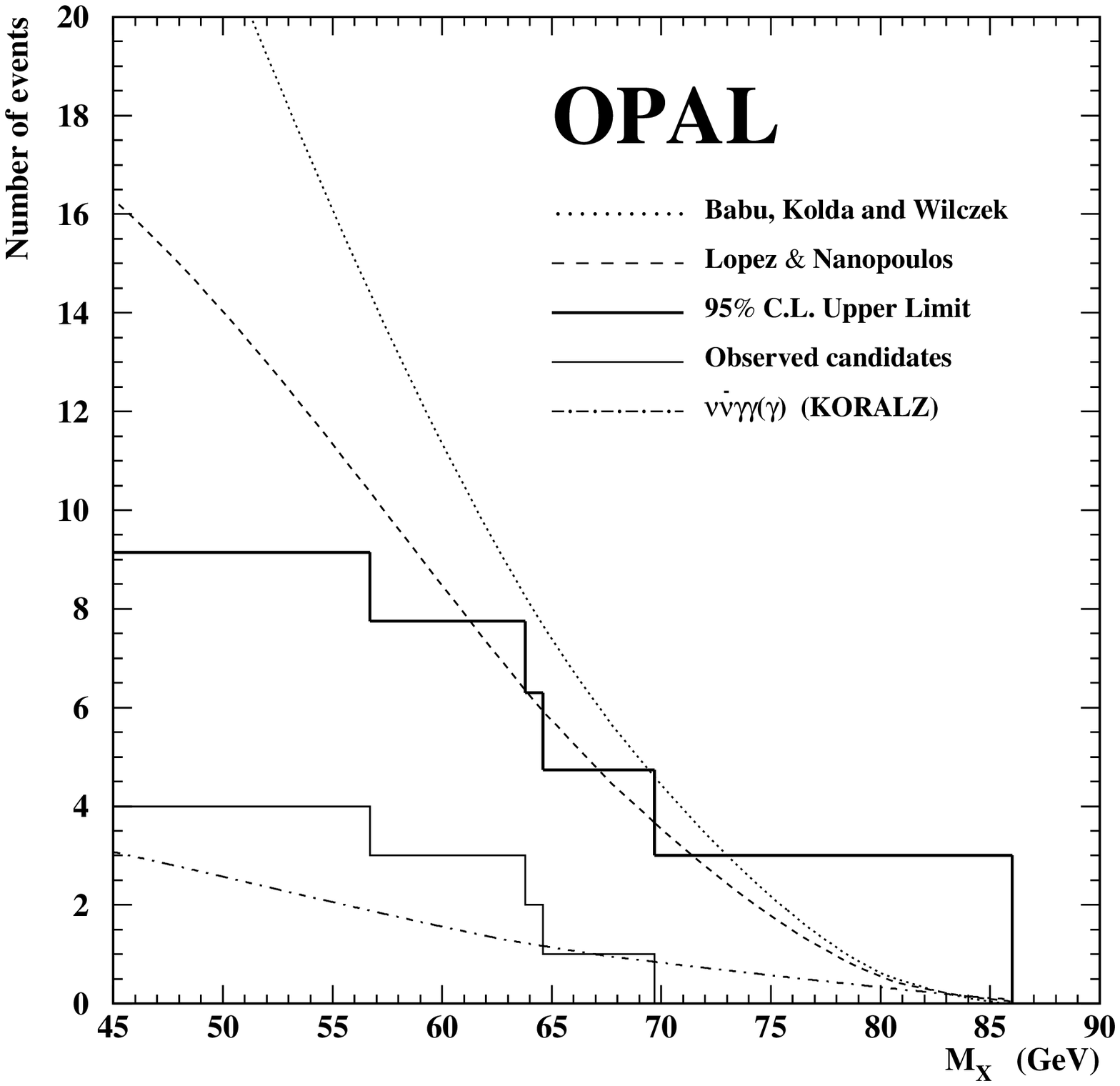,width=11cm,
bbllx=125pt,bblly=150pt,bburx=430pt,bbury=680pt}}
\caption{
Results for 
$\eetoXX$, $\XtoYg$, with $\myzero$
for the combined data sample. The lower solid line shows the number of observed
candidates consistent with a given value of $\mx$. The dashed-dotted
line shows the expected contribution from $\nngggbra$
obtained from the KORALZ generator. 
The thick solid line shows the 95\% confidence level upper limit on the number of
candidate events. The dashed (dotted) line shows 
the expected number of events from the model of 
Lopez and Nanopoulos\cite{gravitinos2} 
(Babu, Kolda and Wilczek\cite{rf:chang}). 
Within these models,
$\lsp$ masses less than 61.3 (69.4) GeV are excluded at the 95\% confidence
level.}
\label{fig:limits_all}
\end{figure}

To set combined, model dependent limits on the mass of the $\lsp$ (NLSP) in
supersymmetric models in which the LSP is a light gravitino, we sum the number of 
observed events consistent with each value of $\mx$.
This distribution is shown in 
Figure~\ref{fig:limits_all} where the solid line shows the number of observed 
candidates consistent with a given value of $\mx$ and the
dashed-dotted line shows the background expectation from KORALZ. 
The number of candidate events is 
consistent with 
the number of background events expected from $\nngggbra$.
The thick solid line shows the 95\% confidence level upper limit at each mass value.
The background expectation is not taken into account when calculating the limit.
Also shown in Figure~\ref{fig:limits_all} are the numbers of events expected from the 
Lopez and Nanopoulos no-scale supergravity model\cite{gravitinos2} and from the 
model of Babu, Kolda and Wilczek\cite{rf:chang} in which 
the  neutralino composition is purely gaugino (bino). For both of these models,
the cross-section has been evaluated at Born level.
Based on these distributions, these two models are ruled out at the 95\% confidence 
level for $M_{\lsp} < 61.3~\GeV$ and $69.4$ GeV, respectively.

As described above, the efficiencies over the full angular range have
been calculated using  isotropic angular distributions for production and 
decay of $\PX$. The validity of this model  has been examined
based on the angular distributions calculated for photino pair 
production in \cite{ELLHAG}. For models proposed in \cite{gravitinos}, the 
production angular distributions are more central and so this procedure is 
conservative. For a $1 + \cos^2{\theta}$ 
production angular distribution, expected for t-channel exchange of a 
very heavy particle according to \cite{ELLHAG}, the 
relative efficiency reduction would be less than 2\% for all
points in the $\mx,\my$ plane (for $\mx-\my > 5$ GeV).
 
Interpretation of these results for the production of excited neutrinos is described
in a separate paper~\cite{ref:opal_exlp_172}.

\section{Conclusions}
We have searched for photonic events with large missing energy in two different and 
complementary topologies in data taken with the OPAL detector at LEP, 
at centre-of-mass energies in the region of 130-172 GeV.

%
%
In the single-photon selection which requires at least one photon with
$x_{T} > 0.05 $ in the region $\acosthe < 0.82$ or one photon
with $x_{T} > 0.1 $ in the region  $15^{\circ}<\theta<165^{\circ}$ 
($\acosthe < 0.966)$
a total of 138 events are observed in the data compared to the KORALZ 
prediction for the contribution from the Standard Model process $\eetonnggbra$
of $141.1 \pm 1.1$ events and an expected
non-physics background of $2.3 \pm 1.1$ events.
The corresponding cross-sections for $\eetonnggbra$
are $10.0 \pm 2.3$, $16.3 \pm 2.8$, $5.3 \pm 0.8$ and $5.5 \pm 0.8$ pb for $\roots$ =
130, 136, 161 and 172 GeV, respectively, in agreement with
the Standard Model expectations. We derive upper limits on the 
cross-section times branching ratio for the process  
$\eetoXY$, $\XtoYg$ for the general case of
massive $\PX$ and $\PY$. The limits vary from 0.31 to 1.8 pb in the
region of interest of the $(\mx,\my)$ plane and include 
the special case of $\myzero$, where the limit varies between
0.36 and 0.76 pb for the $\mx$ mass range from $M_{\rm Z}$
to 172 GeV.

The acoplanar-photons selection requires at least two photons with scaled energy
$x_{\gamma}>0.05$ within the polar angle region $15^{\circ}<{\theta}<165^{\circ}$
or at least two photons with energy $E_{\gamma}>1.75$ GeV with one satisfying
$\acosthe < 0.8$ and the other satisfying $15^{\circ}<{\theta}<165^{\circ}$.
From the combined  data sample 
11 events are selected. The KORALZ prediction for the number of events from 
$\eetonngggbra$
is $6.3\pm{0.2}$. 
The cross-section for this process is measured at 
each centre-of-mass energy (see Table~\ref{tab:g2_nevt}).
Due to the uncertainties in the current modelling of the Standard Model background,
$\eetonngggbra$,  all limits from the acoplanar-photons analysis were
calculated without taking into account the background estimate.
Based on a variety of kinematic distributions, all observed events appear 
consistent with 
$\eetonngggbra$. We derive 95\% CL upper limits on 
$\sigbrXX$ ranging from 0.18 to 0.80 pb for the general case of massive $\PX$ and $\PY$, 
and from 0.35 to 0.50 pb for the special case of $\myzero$.

For the single-photon and acoplanar-photons search topologies, 
the general case of massive $\PX$ and $\PY$ is relevant to the supersymmetry models
in which $\PX = \nln$ and $\PY = \lsp$, with $\XtoYgs$ and $\lsp$ stable.
The special case of $\myzero$ is of particular interest for  
single and pair production of excited neutrinos and for
supersymmetric models in which the LSP is a light gravitino and the $\lsp$ is
the NLSP which decays as $\XtoYgg$.
For the latter scenario, the results of the acoplanar-photons search are used to place
model-dependent lower limits on the $\lsp$ mass.
Comparison with the model predictions of Lopez and Nanopoulos\cite{gravitinos2} 
permits exclusion of that model for $M_{\lsp} < 61.3$ GeV.
A similar model from Babu, Kolda and Wilczek\cite{rf:chang} is excluded for 
$M_{\lsp}<69.4$ GeV. The results of these searches have also been used to 
place limits on the production of excited neutrinos\cite{ref:opal_exlp_172}.

\section{Acknowledgements}

The authors wish to thank J.L. Lopez for providing cross-section results, and
S. Ambrosanio, G.Kane, G. Montagna and Z. W\c{a}s for useful 
discussions on the estimation of the Standard Model cross-sections.

We particularly wish to thank the SL Division for the efficient operation
of the LEP accelerator at all energies
 and for
their continuing close cooperation with
our experimental group.  We thank our colleagues from CEA, DAPNIA/SPP,
CE-Saclay for their efforts over the years on the time-of-flight and trigger
systems which we continue to use.  In addition to the support staff at our own
institutions we are pleased to acknowledge the  \\
Department of Energy, USA, \\
National Science Foundation, USA, \\
Particle Physics and Astronomy Research Council, UK, \\
Natural Sciences and Engineering Research Council, Canada, \\
Israel Science Foundation, administered by the Israel
Academy of Science and Humanities, \\
Minerva Gesellschaft, \\
Benoziyo Center for High Energy Physics,\\
Japanese Ministry of Education, Science and Culture (the
Monbusho) and a grant under the Monbusho International
Science Research Program,\\
German Israeli Bi-national Science Foundation (GIF), \\
Bundesministerium f\"ur Bildung, Wissenschaft,
Forschung und Technologie, Germany, \\
National Research Council of Canada, \\
Research Corporation, USA,\\
Hungarian Foundation for Scientific Research, OTKA T-016660, 
T023793 and OTKA F-023259.\\
%

\newpage
%
%
%
\begin{table}[b]
\centering
\begin{tabular}{|c||c|c|c|c|c|c|c|} \hline
$\roots$(GeV) & 
$\mathrm N_{obs}$ &
${\mathrm N}_{K}^{\nnggbra}$ & 
${\mathrm N}_{N}^{\nnggbra}$ & 
$\mathrm N_{bkg}$ &
${\epsilon}_{K}^{\nnggbra}$(\%) & 
${\epsilon}_{N}^{\nnggbra}$(\%) & 
$\mathrm {\sigma}_{meas}^{\nnggbra}$(pb) 
 \\ \hline \hline
130 & 19 & 25.3 $\pm$ 0.4 & 27.5 $\pm$ 0.3 &
  0.2 $\pm$ 0.2  & 81.6 $\pm$ 0.6 & 79.6 $\pm$ 0.5 & 10.0 $\pm$ 2.3  \\ \hline 
136 &  34 & 23.3 $\pm$ 0.4 & 25.8 $\pm$ 0.3 &
  0.3 $\pm$ 0.2  & 79.7 $\pm$ 0.7 & 80.0 $\pm$ 0.5 & 16.3 $\pm$ 2.8  \\ \hline 
161 & 40 & 48.3 $\pm$ 0.6 & 51.3 $\pm$ 0.4 &
  0.9 $\pm$ 0.8  & 75.2 $\pm$ 0.5 & 72.3 $\pm$ 0.4 & \phantom{1}5.3 $\pm$ 0.8 
 \\ \hline 
172 & 45 & 44.3 $\pm$ 0.6 & 46.1 $\pm$ 0.4 &
  0.9 $\pm$ 0.8  & 77.9 $\pm$ 0.5 & 74.5 $\pm$ 0.4 & \phantom{1}5.5 $\pm$ 0.8 
 \\ \hline \cline{1-5}
130-172 & 138 & 141.2 $\pm$ 1.1 & 150.7 $\pm$ 0.7 &
  2.3 $\pm$ 1.1  
 \\ \cline{1-5}
\end{tabular}
\caption{For each centre-of-mass energy, the table shows the number of events
observed in the OPAL data, the number expected based on the KORALZ ($K$) and
NUNUGPV/NNGG03 ($N$) $\eetonnggbra$ event generators and the number of events expected
from non-physics backgrounds. Also shown are the efficiencies obtained from the
two generators, within the kinematic acceptance of the single-photon selection,
and the measured cross-sections within the kinematic acceptance, determined 
using the efficiencies obtained with the KORALZ generator.
The quoted cross-section errors are statistical.
}
\label{tab:onep_results}
\end{table}
%
%
%
\begin{table}
\centering
\begin{tabular}{|c||c|c|c|c|} \hline
$\mx + \my $ & $\my = 0$ &
$\my = \mx/2$  & $\my = \mx - 15$ GeV & $\my = \mx - 5$ GeV 
\\ \hline \hline
110 & 80.2 $\pm$ 1.7~ & 83.7 $\pm$ 1.5~ & 80.9 $\pm$ 1.7~ &
 51.3 $\pm$ 2.2~ \\ \hline 
120 & 82.3 $\pm$ 1.6~ & 86.6 $\pm$ 1.4~ & 80.2 $\pm$ 1.7~ &
 49.2 $\pm$ 2.2~ \\ \hline 
128 & 84.9 $\pm$ 1.5~ & 84.4 $\pm$ 1.5~ & 82.5 $\pm$ 1.6~ &
 49.6 $\pm$ 2.2~ \\ \hline 
\end{tabular}
\caption{Single photon selection efficiency (\%) as a function of the sum of
$\mx$ and $\my$ versus various $\my$ values
for the process $\eetoXY$, $\XtoYg$.
These efficiencies are for $\roots = 130$ GeV. Masses given are in GeV. }
\label{tab:g1_eff130}
\end{table}
\begin{table}
\centering
\begin{tabular}{|c||c|c|c|c|} \hline
$\mx + \my $ & $\my = 0$ &
$\my = \mx/2$  & $\my = \mx - 15$ GeV & $\my = \mx - 5$ GeV 
\\ \hline \hline
110 & 23.4 $\pm$ 1.3~ & 83.5 $\pm$ 1.1~ & 75.8 $\pm$ 1.3~ &
 38.5 $\pm$ 1.5~ \\ \hline 
130 & 32.1 $\pm$ 1.4~ & 86.4 $\pm$ 1.0~ & 74.6 $\pm$ 1.3~ &
 36.5 $\pm$ 1.5~ \\ \hline 
150 & 65.6 $\pm$ 1.4~ & 85.0 $\pm$ 1.0~ & 75.3 $\pm$ 1.3~ &
 32.2 $\pm$ 1.5~ \\ \hline 
160 & 80.0 $\pm$ 1.2~ & 86.5 $\pm$ 1.0~ & 74.0 $\pm$ 1.3~ &
 30.9 $\pm$ 1.4~ \\ \hline 
170 & 82.9 $\pm$ 1.1~ & 88.7 $\pm$ 0.9~ & 74.8 $\pm$ 1.3~ &
 30.8 $\pm$ 1.4~ \\ \hline 
\end{tabular}
\caption{Single photon selection efficiency (\%) as a function of the sum of
$\mx$ and $\my$ versus various $\my$ values
for the process $\eetoXY$, $\XtoYg$.
These efficiencies are for $\roots = 172$ GeV. Masses given are in GeV. 
The errors are statistical.}
\label{tab:g1_eff172}
\end{table}
%
%
\begin{table}[b]
\centering
\begin{tabular}{|c||c|c|c|c|c|} \hline
$\roots$(GeV) & $\cal{L}$(pb$^{-1}$) & $\mathrm N_{obs}$ &
${\mathrm N}^{\nngggbra}_{\rm exp}$ & 
$\sigma^{\nngggbra}$(pb) & $\sigma^{\nngggbra}_{\rm exp}$(pb) \\ \hline
130 & 2.30  & 3 & $0.83\pm{0.08}$ & $2.0\pm{1.1}$ & $0.54\pm{0.04}$  \\ \hline 
136 & 2.59  & 5 & $0.77\pm{0.08}$ & $3.0\pm{1.3}$   & $0.48\pm{0.04}$  \\ \hline 
161 & 9.89  & 1 & $2.41\pm{0.15}$ & $0.16\pm{0.16}$ & $0.35\pm{0.02}$  \\ \hline 
172 & 10.28 & 2 & $2.28\pm{0.15}$ & $0.30\pm{0.21}$ & $0.33\pm{0.02}$  \\ \hline \cline{1-4}
130-172 & 25.06 & 11 & $6.29\pm{0.24}$ \\ \cline{1-4}
\end{tabular}
\caption{Number of acoplanar-photons events observed and expected 
at each centre-of-mass energy region and the corresponding measured and
expected (KORALZ) cross-sections
within the kinematic acceptance of the acoplanar-photons event selection. 
}
\label{tab:g2_nevt}
\end{table}
%
%
%
\begin{table}
\centering
\begin{tabular}{|c||c|c|c|c|c|c|c|c|c|} \hline
$\roots$(GeV) & $x_1$ & $x_2$  & $\mathrm cos{\theta}_1$ & 
$\mathrm cos{\theta}_2$ & ${\phi}_1(\rm rad)$ & $\phi_2(\rm rad)$  & $M_{\rm recoil}$ 
& $M_{\gamma\gamma}$ & $\mxmax$  \\ \hline \hline
130.3 & 0.313 & 0.048 &  0.785 & -0.721 & 5.134 & 0.095 & 105.0  & 13.5 & 20.6 \\ \hline 
130.3 & 0.435 & 0.276 &  0.473 & -0.926 & 4.064 & 1.172 &  81.9  & 42.3 & 51.7 \\ \hline 
130.3 & 0.449 & 0.091 &  0.484 &  0.166 & 2.147 & 4.574 &  91.4  & 23.3 & 25.6 \\ \hline 

136.2 & 0.499 & 0.069 & -0.800 &  0.024 & 0.377 & 2.642 &  92.0  & 21.2 & 24.8 \\ \hline 
136.2 & 0.529 & 0.036 & -0.154 & -0.506 & 4.105 & 4.799 &  90.2  &  6.8 & 24.8 \\ \hline 
136.2 & 0.456 & 0.150 &  0.420 &  0.896 & 0.692 & 3.408 &  91.8  & 33.2 & 32.1 \\ \hline 
136.2 & 0.512 & 0.031 & -0.231 & -0.026 & 5.845 & 4.147 &  93.0  & 12.9 & 18.2 \\ \hline 
136.2 & 0.515 & 0.070 &  0.465 &  0.413 & 1.413 & 2.484 &  88.5  & 11.9 & 34.0 \\ \hline 

161.3 & 0.402 & 0.166 &  0.580 &  0.095 & 1.206 & 0.581 & 107.2  & 15.8 & 59.6 \\ \hline 
172.3 & 0.592 & 0.170 &  0.046 &  0.787 & 5.509 & 3.776 &  93.0  & 39.9 & 58.8 \\ \hline 
172.3 & 0.565 & 0.209 &  0.901 & -0.269 & 1.510 & 2.866 &  93.5  & 44.9 & 64.7 \\ \hline 
\end{tabular}
\caption{Kinematic properties of the events passing the acoplanar-photons
selection. All masses are given in GeV. $\mxmax$
is defined in Section~\ref{sec:g2_results_my0}.
}
\label{tab:g2_kine}
\end{table}
\begin{table}
\centering
\begin{tabular}{|c||c|c|c|c|}
\hline
$\mx$ & $\my$=0 & $\my=\mx /2$ & $\my=\mx-15$ GeV & $\my=\mx-5$ GeV \\ \hline \hline
85 & $72.6\pm{1.3}$ & $71.0\pm{1.3}$ & $69.1\pm{1.4}$ & $43.2\pm{1.5}$ \\ \hline
80 & $72.6\pm{1.3}$ & $73.3\pm{1.3}$ & $71.1\pm{1.3}$ & $41.2\pm{1.5}$  \\ \hline
75 & $73.6\pm{1.3}$ & $72.6\pm{1.3}$ & $69.7\pm{1.4}$ & $39.9\pm{1.4}$  \\ \hline
70 & $71.8\pm{1.3}$ & $69.2\pm{1.4}$ & $68.9\pm{1.4}$ & $42.3\pm{1.5}$  \\ \hline
55 & $68.1\pm{1.4}$ & $67.7\pm{1.4}$ & $65.8\pm{1.4}$ & $41.8\pm{1.5}$  \\ \hline
45 & $67.3\pm{1.4}$ & $66.1\pm{1.4}$ & $64.2\pm{1.4}$ & $40.6\pm{1.4}$  \\ \hline
\end{tabular}
\caption[]{Acoplanar-photons selection efficiencies (\%) 
for the process $\eetoXX$, $\XtoYg$ at $\roots = 172$ GeV for various
$\mx$ and $\my$ (in GeV). These values include the efficiency of
the kinematic consistency cuts. Similar efficiencies are obtained at 
each value of $\roots$. The errors are statistical.
}
\label{tab:g2_eff172}
\end{table}
\begin{table}
\centering
\begin{tabular}{|c||c|c|c|}
\hline
  & Selection efficiency for  &
Selection efficiency with & $\nngggbra$ rejection \\
 $\mx$ & $\eetoXX$, $\XtoYg$ & 
 $\mxmax>\mx-5$ GeV & 
efficiency \\ \hline
\hline
85 & $74.7 \pm 1.2$  & $72.5 \pm 1.2$  & $95.7 \pm 1.3$  \\ \hline
80 & $74.5 \pm 1.2$  & $70.7 \pm 1.3$  & $89.7 \pm 2.0$  \\ \hline
75 & $74.9 \pm 1.2$  & $72.7 \pm 1.2$  & $85.4 \pm 2.3$  \\ \hline
70 & $72.7 \pm 1.2$  & $69.6 \pm 1.3$  & $79.4 \pm 2.7$  \\ \hline
55 & $68.7 \pm 1.3$  & $66.1 \pm 1.3$  & $58.4 \pm 3.2$  \\ \hline
45 & $68.1 \pm 1.3$  & $64.8 \pm 1.4$  & $41.2 \pm 3.2$  \\ \hline
\end{tabular}
\caption[]{
Acoplanar-photons event selection efficiency (\%), as a function of mass, 
for the process $\eetoXX$, $\XtoYg$, for $\myzero$. These numbers 
are for $\roots = 172$ GeV. The first column shows the efficiency  
of the selection described in section~3.2. The second column shows the selection efficiency after the
cut on $\mxmax$ described in section~4.2.2.
The final column shows the rejection efficiency (\%) of the 
$\mxmax>\mx-5$ GeV cut for $\nngggbra$ events. The errors are statistical. 
}
\label{tab:g2_eff172_my0}
\end{table}
%
%
\clearpage
\newpage
\clearpage
\newpage
\newpage
%
\newpage
\newpage
\newpage
\newpage
\newpage
\newpage
\newpage
\newpage
\newpage
\newpage
\newpage
\clearpage
\newpage
\newpage
\newpage
\newpage

\begin{thebibliography}{99}
\bibitem{rf:OPALSP130}
  OPAL Collab., G. Alexander et al.,
  Phys. Lett. {\bf B377} (1996) 222.
\bibitem{rf:OPALSP161} OPAL Collab., K. Ackerstaff et al.,
  Phys. Lett. {\bf B391} (1997) 210.
\bibitem{rf:excitedl}
  OPAL Collab., G. Alexander et al.,
  Phys. Lett. {\bf B386} (1996) 463.
\bibitem{rf:excited161}
  OPAL Collab., K. Ackerstaff et al.,
  Phys. Lett. {\bf B391} (1997) 197.
\bibitem{OPALgg}
  OPAL Collab., K. Ackerstaff et al.,
  CERN-PPE/97-109, submitted to Zeit. Phys. {\bf C}. 
\bibitem{rf:OPALSP94}
  OPAL Collab., R. Akers et al.,
  Z. Phys. {\bf C65} (1995) 47.
\bibitem{rf:LEPSP}
  L3 Collab., B. Adeva et al.,
  Phys. Lett. {\bf B275} (1992) 209;\newline
  L3 Collab., O. Adriani et al.,
  Phys. Lett. {\bf B292} (1992) 463;\newline
  ALEPH Collab., D. Buskulic et al.,
  Phys. Lett. {\bf B313} (1993) 520;\newline
  DELPHI Collab., P. Abreu et al.,
  Z. Phys. {\bf C74} (1997) 577.
\bibitem{rf:lowe}
  MAC Collab., W.T. Ford et al., 
  Phys. Rev. {\bf D33} (1986) 3472;\newline 
  H. Wu, Ph.D Thesis, Univ. Hamburg, 1986;\newline
  CELLO Collab., H.-J. Behrend et al., 
  Phys. Lett. {\bf B215} (1988) 186;\newline
  ASP Collab., C. Hearty et al.,
  Phys. Rev. {\bf D39} (1989) 3207;\newline
  VENUS Collab., K. Abe et al., 
  Phys. Lett. {\bf B232} (1989) 431;\newline
  TOPAZ Collab., T.~Abe et al.,
  Phys. Lett. {\bf B361} (1995) 199.
\bibitem{rf:LEPSP130}
  ALEPH Collab., D. Buskulic et al.,
  Phys. Lett. {\bf B384} (1996) 333;\newline
  L3 Collab., M. Acciarri et al.,
  Phys. Lett. {\bf B384} (1996) 323;\newline
  DELPHI Collab., P. Abreu et al., 
  Phys. Lett. {\bf B380} (1996) 471;\newline
  L3 Collab., M. Acciarri et al.,
  CERN-PPE/97-76, submitted to Phys. Lett. {\bf B};\newline
  ALEPH Collab., R.Barate et al.,
  CERN-PPE/97-122, submitted to Phys. Lett. {\bf B}.
%
\bibitem{Kane}
S. Ambrosanio et al., Phys. Rev. Lett. {\bf 76} (1996) 3498;
Phys. Rev. {\bf D55} (1997) 1392.
\bibitem{radN2} 
H.E. Haber and D. Wyler,
Nucl.\ Phys.\ {\bf B323} (1989) 267;\newline
S. Ambrosanio and B. Mele, Phys. Rev. {\bf D53} (1996) 2541.
\bibitem{ELLHAG}
J. Ellis and J.S. Hagelin, Phys. Lett. {\bf B122} (1983) 303.
\bibitem{gravitinos}
S. Dimopoulos et al., Phys. Rev. Lett. {\bf 76} (1996) 3494;\newline
D.R. Stump, M. Wiest, C.P. Yuan, Phys. Rev. {\bf D54} (1996) 1936;\newline
S. Ambrosanio et al., Phys. Rev. {\bf D54} (1996) 5395.
%
\bibitem{gravitinos2}
J.L. Lopez and D.V. Nanopoulos, Mod. Phys. Lett. {\bf A11} (1996) 2473;
Phys. Rev. {\bf D55} (1997) 4450.
\bibitem{LNZ}
J.L. Lopez, D.V. Nanopoulos, A. Zichichi, Phys. Rev. Lett. {\bf 77} (1996) 5168.
\bibitem{rf:chang}
C.Y. Chang and G.A. Snow, UMD/PP/97-57. \\
K. S. Babu, C. Kolda and F. Wilczek, Phys. Rev. Lett. {\bf 77} (1996) 3070.
\bibitem{OPAL_Hgg}
OPAL Collab, K.~Ackerstaff et al., 
CERN-PPE/97-121, submitted to Z. Phys. {\bf C}.
%
\bibitem{rf:OPAL-detector}
  \OPALColl, K.~Ahmet et~al., \NIM\ {\bf A305} (1991) 275;\newline
  P.P.~Allport et~al., \NIM\  {\bf A324} (1993) 34;\newline
  P.P.~Allport et~al., \NIM\  {\bf A346} (1994) 476;\newline
  B.E.~Anderson et~al., IEEE Transactions on Nuclear Science {\bf 41}
  (1994) 845.
\bibitem{trigger}
M. Arignon et al., \NIM\ {\bf A313} (1992) 103.
\bibitem{rf:vvgmc}
  F.A.~Berends et al., Nucl.\ Phys.\ {\bf B301} (1988) 583;\newline
  R.~Miquel, C.\ Mana and M.\ Martinez,
  Z.\ Phys.\ {\bf C48} (1990) 309.
\bibitem{rf:paviaMC}
  G. Montagna et al., Nucl. Phys. {\bf B452} (1996) 161; \\
  G. Montagna, O. Nicrosini and F. Piccinini, FNT/T-96/1,
   to be published in Comp. Phys. Comm. 
\bibitem{rf:KORALZ}
  S.~Jadach et al., Comp. Phys. Comm. {\bf 66} (1991) 276.
  Version 4.02 was used including a recommended correction to the 
  NDIST0 subroutine.
%
\bibitem{rf:RADCOR}
  F.A.~Berends and R.~Kleiss,
  Nucl.\ Phys.\ {\bf B186} (1981) 22.   
\bibitem{rf:BHWIDE}
S.~Jadach, W.~Placzek and B.~F.~L.~Ward, 
Phys. Lett. {\bf B390} (1997) 398. 
\bibitem{rf:TEEGG}
  D. Karlen, Nucl.\ Phys.\ {\bf B289} (1987) 23.
\bibitem{rf:SUSYGEN}
S. Katsanevas and S. Melachronios, CERN/96-01, Vol.2 (1996) 328. 
%
\bibitem{rf:GOPAL}
  J.~Allison et~al., \NIM\  {\bf A317} (1992) 47.
\bibitem{rf:colas}
P.~Colas, R.~Miquel and Z.~W\c{a}s, Phys. Lett. {\bf B246} (1990) 541.
\bibitem{rf:YFS}
D.~R.~Yennie, S.~C.~Frautschi and H. Suura, Annals of Phys. {\bf 13} (1961) 379. 
\bibitem{rf:comphep}
E.~E.~Boos et al., ``COMPHEP: Specialized Package for Automatic Calculations of Elementary Particle
Decays and Collisions'', SNUTP-94-116.
\bibitem{rf:Sandro}
S.~Ambrosanio, G.~D.~Kribs and S.~P.~Martin,
Phys. Rev. {\bf D56} (1997) 1761.
\bibitem{rf:helas}
H.~Murayama et al., ``HELAS: Helicity Amplitude Subroutines for Feynman Diagram Evaluations'', 
KEK 91-11 (1992).
\bibitem{rf:Mrenna}
S. Mrenna, ``Physics Backgrounds to Supersymmetric Signals with Two Photons and Missing Mass at LEP'',
ANL-HEP-PR-97-27.
\bibitem{rf:bread-bath}
P. Bain and R. Pain, ``Acoplanar Photons Background Generators'', Contributed paper number 528, 
EPS conference, Jerusalem, 1997.
\bibitem{rf:grace}
MINAMI-TATEYA group, ``GRACE Manual: Automatic Generation of Tree Amplitudes in Standard Models, 
Version 1.0'', KEK report 92-19 (1993).
\bibitem{ref:opal_exlp_172}
OPAL Collab., K. Ackerstaff et al.,
CERN-PPE/97-123, submitted to Z. Phys. {\bf C}.
\bibitem{rf:LEP1XY}
  L3 Collab., M.~Acciarri et al.,
  Phys. Lett. {\bf B350} (1995) 109;\newline
  OPAL Collab., G.~Alexander et al.,
  Phys. Lett. {\bf B377} (1996) 273. 
\bibitem{LEP1XX}
  OPAL Collab., M.Z.~Akrawy et al.,
  Phys. Lett. {\bf B248} (1990) 211;\newline
  ALEPH Collab., D.~Decamp et al., 
  Phys. Rep. 216 (1992) 253;\newline
  L3 Collab., M.~Acciarri et al.,
  Phys. Lett. {\bf B350} (1995) 109.
\bibitem{rf:PDG96} Particle Data Group, R. M. Barnett et al., Phys. Rev. {\bf D54}
(1996) 1.
\bibitem{rf:systerr} R.~D.~Cousins and V.~L.~Highland, 
  \NIM\ {\bf A320} (1992) 331.
\end{thebibliography}
\end{document}